\documentclass[journal]{IEEEtran}

\usepackage[compress]{cite}   %citation order
\usepackage{amsmath, bm}   % math package from American Mathematical Society
\usepackage{amsfonts}      % fonts package from American Mathematical Society
\usepackage{graphicx}
\usepackage{pdfpages}
\usepackage{subfig}   % set "caption=false" package option to prevent overriding IEEEtrans.cls caption
\usepackage{booktabs} %w insert table
\usepackage{cleveref}
\usepackage{multirow}
\usepackage{url}   % cite url

\usepackage{regexpatch,fancyvrb,xparse} % for footnote
\makeatletter
\let\do@footnotetext\@footnotetext
\regexpatchcmd{\do@footnotetext}
  {\c{insert}\c{footins}\cB.(.*)\cE.}
  {\1\c{egroup}}
  {}{}
\def\@footnotetext{\insert\footins\bgroup\@makeother\#\do@footnotetext}
\newcommand{\ttvar}{\begingroup\@makeother\#\@ttvar}
\newcommand{\@ttvar}[1]{\ttfamily\detokenize{#1}\endgroup}
\makeatother  

\crefname{figure}{Fig.}{Figs.} 
\begin{document} % the beginning of documentation, such as main() in C language
	% title and author names
	%\title{Denoising autoencoder based on data synthesis with Spatial-temporal Constraints for Electromagnetic Source Imaging}
%	\title{Enhancing Electromagnetic Source Imaging via a Denoising Autoencoder}
	\title{	Electromagnetic Source Imaging via a Data-Synthesis-Based Convolutional Encoder-Decoder Network}
	\author{Gexin~Huang, Jiawen~Liang,
		Ke~Liu,
		Chang~Cai,		 
		Zheng Hui~Gu,~\IEEEmembership{Member,~IEEE,}
		Feifei~Qi,
		Yuanqing~Li,~\IEEEmembership{Fellow,~IEEE,}
		Zhu Liang~Yu\IEEEauthorrefmark{1},~\IEEEmembership{Member,~IEEE,}
		and~Wei~Wu\IEEEauthorrefmark{1},~\IEEEmembership{Senior Member,~IEEE,}
	\thanks{This work was supported in part by the National Natural Science Foundation of China under Grants 61836003, 61906048, and 61703065.}
	\thanks{Z. L. Yu, Z. Gu, Y. Li, and J. Liang are with the School of Automation Science and Engineering, South China University of Technology, Guangzhou, China, 510641. Wei. Wu is with the Department of Psychiatry and Behavioral Sciences, Stanford University, Stanford, CA, USA. K. Liu is with Chongqing Key Laboratory of Computational Intelligence, Chongqing University of Posts and Telecommunications, Chongqing, 400065, China. C. Cai is with the National Engineering Research Center for E-Learning, Central China Normal University, Wuhan, China. F. Qi is with School of Internet Finance and Information Engineering, Guangdong University of Finance, Guangzhou, 510521, China (* Corresponding author: Wei Wu. E-mail: weiwuneuro@gmail.com, Zhu Liang Yu. E-mail: zlyu@scut.edu.cn)}	
	}
	\markboth{IEEE Transactions on Neural Networks and Learning Systems}%
	{Huang \MakeLowercase{\textit{et al.}}: IEEE Transactions on Neural Networks and Learning Systems}

	\maketitle
	% abstract
	\begin{abstract}
	% modified in 6/10f
	Electromagnetic source imaging (ESI) requires solving a highly ill-posed inverse problem. To seek a unique solution, traditional ESI methods impose various forms of priors that may not accurately reflect the actual source properties, which may hinder their broad applications. To overcome this limitation, in this paper a novel data-synthesized spatio-temporally convolutional encoder-decoder network method termed DST-CedNet is proposed for ESI. DST-CedNet recasts ESI as a machine learning problem, where discriminative learning and latent-space representations are integrated in a convolutional encoder-decoder network (CedNet) to learn a robust mapping from the measured electroencephalography/magnetoencephalography (E/MEG) signals to the brain activity. In particular, by incorporating prior knowledge regarding dynamical brain activities, a novel data synthesis strategy is devised to generate large-scale samples for effectively training CedNet. This stands in contrast to traditional ESI methods where the prior information is often enforced via constraints primarily aimed for mathematical convenience. Extensive numerical experiments as well as analysis of a real MEG and Epilepsy EEG dataset demonstrate that DST-CedNet outperforms several state-of-the-art ESI methods in robustly estimating source signals under a variety of source configurations.

	\end{abstract}
	% keyword
	% intro
	\section{Introduction}\label{sec.introduction}
	% modified in 6/10
	The forward problem of electroencephalography/magnetoencephalography (E/MEG) involves determining the external surface potentials or magnetic fields from a given configuration of neuronal current activities. In contrast, reconstructing sources of brain activities from the external measurements of electromagnetic signals is defined as the E/MEG inverse problem, also known as the electromagnetic source imaging (ESI) problem. The solution to the ESI problem provides comprehensive spatial–temporal patterns of brain activities associated with a multitude of functions, leading to broad applications in the diagnosis and prediction of brain diseases\cite{BinHe2011,Michel2004,Baillet2001}. 
	% modified in 6/10
	However, due to the limited number of channels for the external measurements compared to the number of unidentified sources, the ESI problem is highly ill-posed, i.e., there are infinite possible solutions for a given measured E/MEG signal, which poses a long-standing challenge for ESI.
		
	To enable a unique solution, previous studies were centered on the optimization or probabilistic framework where regularization techniques or pieces of prior information are exploited to constrain the solution space. Minimum norm estimate (MNE)\cite{Hmlinen1994} employs the $L_2$-norm regularization to estimate spatially smooth sources. Its variant, weighted MNE (wMNE)\cite{Dale1993}, assigns additional weights to the sources to mitigate the bias towards superficial sources. Low-resolution electromagnetic tomography (LORETA)\cite{PascualMarqui1994} is an alternative $L_2$-norm-based approach that utilizes the Laplace matrix to spatially regularize the solution. Although these methods may be suitable for imaging spatially extended sources, the $L_2$-norm regularization may lead to overly diffuse estimates for focal sources\cite{Grech2008,Becker2015}. To overcome this limitation, focal underdetermined system solution (FOCUSS)\cite{Gorodnitsky1995} and sparse source imaging (SSI)\cite{ding2008sparse} introduce $L_p$-norm regularizers that encourage sparsity of the solution. Along this line, the $L_p$ norm iterative sparse solution (LPISS)\cite{xu2007lp} is an iterative sparse learning algorithm based on a $L_p$ norm. Sparse Bayesian learning approaches (SBL)\cite{Wipf2009,owen2012performance,wu2015bayesian} cast the inverse problem under a empirical Bayesian framework where hyperparameters can be automatically determined and sparse solutions can be obtained. In particular, Champagne\cite{Wipf2010,cai2020robust} is an SBL approach that estimates the number, location, and time course of the sources in a principled fashion. Nevertheless, these sparse approaches are fundamentally limited in recovering the spatial extents for extended sources and may suffer from discontinuity in the reconstructed time courses under low signal-to-noise settings \cite{liu2015straps}. In addition to the smoothness and sparseness constraints, prior studies also explored the use of other constraints for source imaging. Spatiotemporal tomographic nonnegative independent component analysis (STTONNICA)\cite{ValdsSosa2009} decomposes source signals into temporal and spatial components via nonnegative matrix factorization. In contrast, methods such as\cite{Awan2018, Knsche2013, Dale2000} exploit the physical properties of neural activity, information about the cortical tissue structures, or other modalities with high spatial resolutions such as Magnetic Resonance Imaging (MRI) to obtain more accurate source imaging results.

	% modified in 6/12
	The aforementioned approaches attempt to solve the inverse problem in a direct manner that use varying forms of prior assumptions to constrain the solution space of the optimization problem. From a machine learning perspective, since there is no labeled data involved to train the underlying models, these approaches can also be viewed as unsupervised learning methods. However, prior information regarding source signals is enforced via constraints often adopted to enable mathematical tractability yet may not accurately reflect the actual source configurations. Moreover, computational complexity may be increased considerably with more accurate modeling analysis\cite{arridge2019solving}. For the purpose of learning the mapping from the sensor to the source space, some prior studies employed artificial neural networks to solve the dipole fitting problem \cite{abeyratne1991artificial,van2000eeg,yuasa1998eeg,abeyratne2001eeg}, which assumes a small number of dipole sources are activated. Nonetheless, the problem is tractable only for one or two dipoles due to the difficulty of optimizing the neural network with more dipoles. Thanks to the rapid development of deep learning in recent years \cite{goodfellow2016deep}, deep neural networks are now capable of learning sophisticated mappings through large-scale training sets for distributed ESI, in which the mappings are represented by stacked nonlinear layers. Thus, deep neural networks may serve as a potential new solution for the ESI problem. Recently, the source imaging framework network (SIFNet)\cite{Sun2020sifnet} was proposed in an attempt to convert the ESI inverse problem to a multi-classification problem, which falls in the realm of supervised learning. 
	% modified in 6/12
	Specifically, after modeling the source space as a series of interconnected regions, a residual-block-based classification network can be successfully trained via the generated training set, where a binary training label is determined at each region to indicate its state (active or inactive). As such, SIFNet is not able to yield estimates of source time series, which nonetheless may provide critical information in many basic or translational studies. Edge Sparse Basis Network (ESBN)\cite{wei2021edge} treats source activities as independently and identically distributed (IID) samples and builds a regression model to estimate the amplitude of source separately at each time point. Accordingly, its data synthesis stage generates IID samples whose amplitudes are randomly sampled from Gaussian distribution and dose not simulate data with temporal dynamics.  
	
	% modified in 6/13
	In this paper, we propose a novel data-synthesized spatio-temporally convolutional encoder-decoder network (DST-CedNet) method that reformulates the ESI problem as a machine learning problem, where discriminative learning and latent-space representations are integrated in a convolutional encoder-decoder 
% 	denoising autoencoder \cite{vincent2010stacked} 
	to learn a robust mapping from E/MEG signals to sources. Notably, by incorporating the spatio-temporal properties of dynamical brain activities, a data synthesis strategy is devised to generate large-scale training samples for effectively training the CedNet, where no complicated mathematical modeling of prior information is required as in the traditional framework. In addition, the structure of the CedNet is tailored to E/MEG signals in that there are dedicated layers performing spatio-temporal filtering of E/MEG signals, ensuring that the spatio-temporal patterns of E/MEG signals are sufficiently leveraged to enhance the performance. 
	
	Compared to SIFNet which focuses on classifying the activation state of each brain region, DST-CedNet is able to not only locate the region of brain activity but also effectively reconstruct their time courses. 
	\textbf{Furthermore, the latent-representation learning task introduced in CedNet renders DST-CedNet robust to the noise in the E/MEG signals.} 
	Experimental results show that the network can recover the spatial extents and the time courses of the sources under even low signal-to-noise settings, outperforming existing ESI algorithms.

	The remainder of the paper is organized as follows. The proposed DST-CedNet approach is introduced in Section \ref{sec.methodology}. Section \ref{sec.experiments} presents the experimental results on simulated and real datasets. Discussions about the merits and limitations of DST-CedNet are presented in Section \ref{sec.discussion}. Finally, conclusions are provided in Section \ref{sec.conclusion}.
	
	% method
	\section{Methodology}\label{sec.methodology}
	
	\subsection{Problem Formulation} \label{sec.methodology.DM}

	The forward model of the E/MEG signals can be represented as
	\begin{equation}
	\label{eqn1}
	\bm{X} = \bm{L} \bm{S} + \bm{\varepsilon},
	\end{equation}
	where $\bm{X}=[\bm{x}_1,\cdots,\bm{x}_t,\cdots,\bm{x}_{T}] \in \mathbb{R}^{{ N_c }\times T}$ denotes the measured E/MEG signals of $N_c$ channels at $T$ time points. Each column $\bm{x}_t = [x_{1t},\cdots,x_{N_c t}]^\mathsf{T}$ represents a snapshot of the signal vector at time $t$. $\bm{S} = [\bm{s}_1,\cdots,\bm{s}_t,\cdots,\bm{s}_{T}] \in \mathbb{R}^{N_s\times T}$ denotes the source signals. Each column $\bm{s}_t = [s_{1t},\cdots,s_{N_s t}]^\mathsf{T} $ represents the source signal vector at time $t$. $\bm{\varepsilon} \in \mathbb{R}^{N_c\times T}$ denotes the measurement noise. $\bm{L}$ is the the lead field matrix, which encodes how each unit source at a certain location is related to the measured E/MEG signals. The lead field matrix can typically be obtained by modeling tissues of the human head via numerical methods such as the finite element model (FEM)\cite{He1987} or boundary element model (BEM)\cite{Hamalainen1989}, based on structural MRI scans of individual subjects or an average template.
	%Yan1991,  Wolters2006,
	
Assuming the lead field matrix $\bm{L}$ has been obtained by solving the forward problem, the ESI problem aims to find a mapping $P(\cdot)$ from E/MEG signals $\bm{X}$ to source signals $\bm{S}$:
	\begin{equation}
	\label{neq2}
	\bm{X} \xrightarrow{P(\cdot)} \bm{S}.
	\end{equation}
	% 接一句，由于XXX，所以这是一个欠定的问题
	% 欠定问题通常是通过叠加线性的正则化项来求解。

	Due to the ill-posedness of the ESI problem when $N_s \gg N_c$, traditional methods introduce regularization techniques in order to obtain a unique solution $\bm{S}^\star$. This can be typically performed explicitly by adding a regularization term to the loss function under the optimization framework or implicitly via priors in probabilistic modeling, both leading to the following solution:
	
	\begin{equation}
	\label{neq3}
	\bm{S}^\star =  \arg \min_{\bm{S}} D(\bm{S})+R(\bm{S}),	
	\end{equation}
	%	\bm{S}^\star =  \arg \min_{\bm{S}} D(\bm{X},\bm{S})+R(\bm{S}),	
where $D(\bm{S})$ measures the discrepancy between the actual and the reconstructed sensor signals based on the estimated source signals, e.g., $D(\bm{S})=\|\bm{X}-\bm{LS}\|_\mathsf{F}^2$ \cite{Hmlinen1994,Dale1993}, where $\|\cdot\|_\mathsf{F}$ denotes the Frobenius norm. $R(\bm{S})$ is the regularization term based on the estimated source signals. For instance, $R(\bm{S})$ can be employed to promote the sparsity ($L_1$ norm) or minimize the overall energy ($L_2$ norm) of the source signals \cite{Awan2018}.

	To overcome the limitation that $R(\bm{S})$ is often devised for mathematical convenience yet may not accurately reflect the actual source configurations, in this paper we propose a novel deep learning approach based on a data synthesis strategy.

	\subsection{Data-Synthesized Deep Learning Strategy}
	\label{sec.methodology.GI}

DST-CedNet views the ill-posed inverse problem as a multi-target regression problem that aims to uncover the mapping $P_{\theta} (\cdot)$ via a nonlinear neural network, where ${\theta}$ represents the unknown weights of the neural network. 
	% modified in 6/14
In order to render the neural network approach feasible, sufficient source and E/MEG data samples should be provided for training the neural network. However, in vivo source data samples are hardly acquired in practice as direct recordings of electrical brain activities are invasive. 

	% modified in 6/16
In this paper, to better reflect the spatio-temporal properties of the source signals, we inject prior information regarding the temporal realistic activities and the spatial compactness into data synthesis. These properties encoded in the synthesized data can then be effectively learned by training the neural network, of which the specific architecture is illustrated in Section \ref{sec.methodology.DD}.
	% the smoothness of their time courses and the anatomy of the head tissues

\begin{figure}[htbp]
		\centering
		\includegraphics[width=0.475\textwidth]{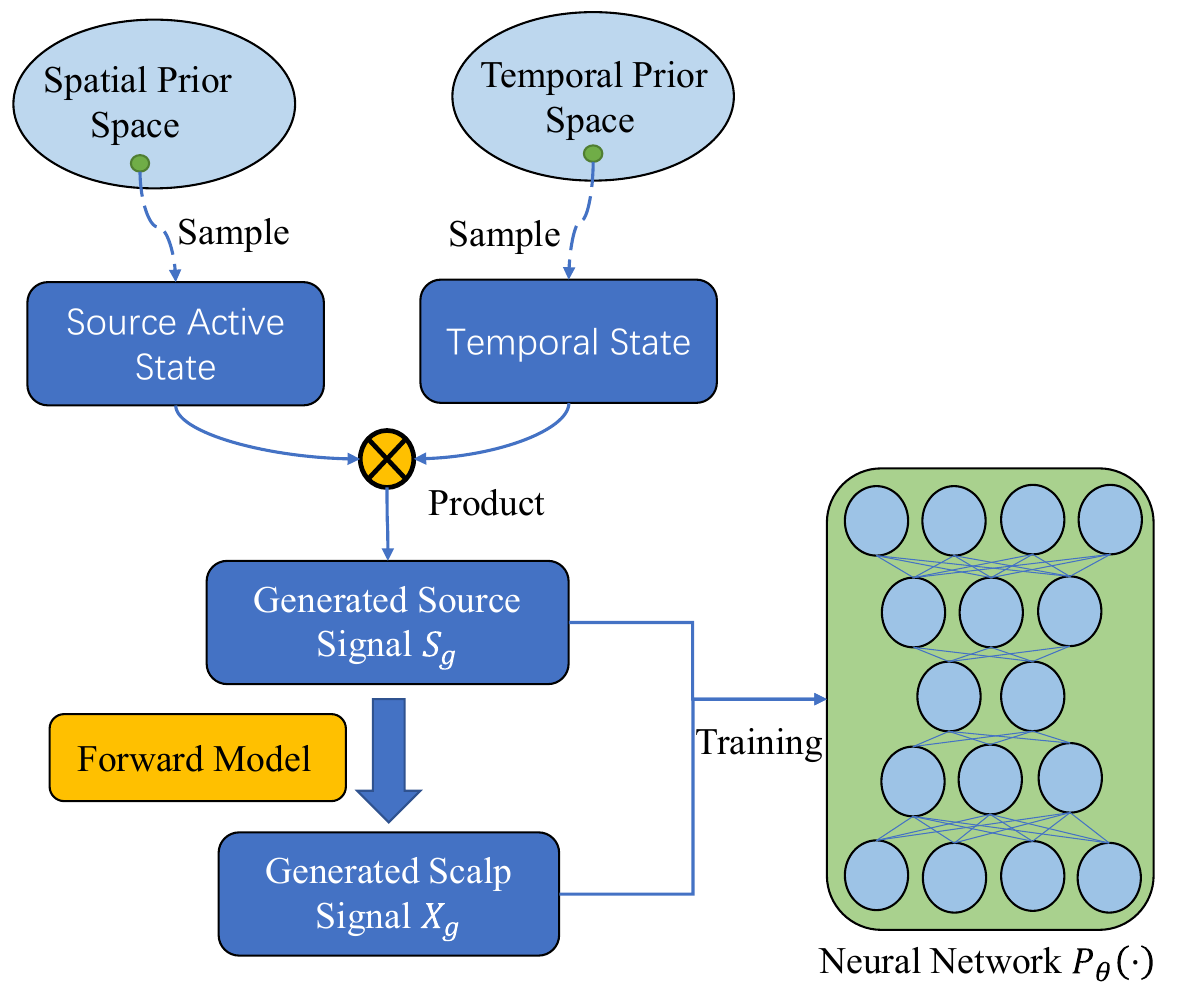}
		\caption{Data-synthesized deep learning strategy for solving the ESI problem.}
		%	\hfil
		\label{fig_1}
	\end{figure}

	The proposed DST-CedNet is depicted in Fig. \ref{fig_1}. Specifically,
	the data synthesis strategy is based on the realistic forward model derived from anatomical MRI scans and temporal basis functions derived from realistic scalp signals.
	% modified in 6/16
	First, the source data are decomposed into temporal and spatial components where various spatial and temporal states are generated by 
	uniformly sampling from the respective prior space (Details are provided in section \ref{sec.methodology.IP}).
	Second, the generated source signals, denoted as $\bm{S}_g$, are obtained via the product of the temporal signals and the source active states, and the corresponding E/MEG signals, denoted as $\bm{X}_g$, can be simulated by passing $\bm{S}_g$ through the forward model. 
	As such, a large-scale dataset of source and E/MEG signals with varying states can be generated as the training set to learn $P_{\theta} (\cdot)$. Under the loss function $\mathcal{L}_{\theta}(\cdot)$, the optimal parameters can be obtained as $\theta^\star=\arg \min_{\theta} \mathcal{L}_{\theta}(\bm{X}_g,\bm{S}_g)$ based on the training set. Finally, the learned inverse mapping $P_{\theta^\star}(\cdot)$ can be used for ESI on unseen E/MEG signals. The implementation details are provided in Section \ref{sec.methodology.IP}.
	
	\subsection{Neural Network Design} \label{sec.methodology.DD}

	As opposed to taking a pure discriminative approach to solving the multi-target regression problem, which may be prone to overfitting in the presence of noise in the E/MEG signals, we propose a more robust approach that integrates discriminative learning and latent-space representation under the framework of the CedNet. Inspired by the EEG decoding network \cite{schirrmeister1703deep, 8310961,8897723}, through blocks with spatio-temporal stepwise convolution/deconvolution tailored to E/MEG signals, 
	%update
	\textbf{the latent-space representation learning in the CedNet is capable of extracting robust features from noisy E/MEG signals that capture the underlying dependencies and spatio-temporal structure, hence effectively regularizing the solution when combined with discriminative learning.}

        \subsubsection{Architecture}	
\begin{figure}[htbp]
		\centering
		\includegraphics[width=0.475\textwidth]{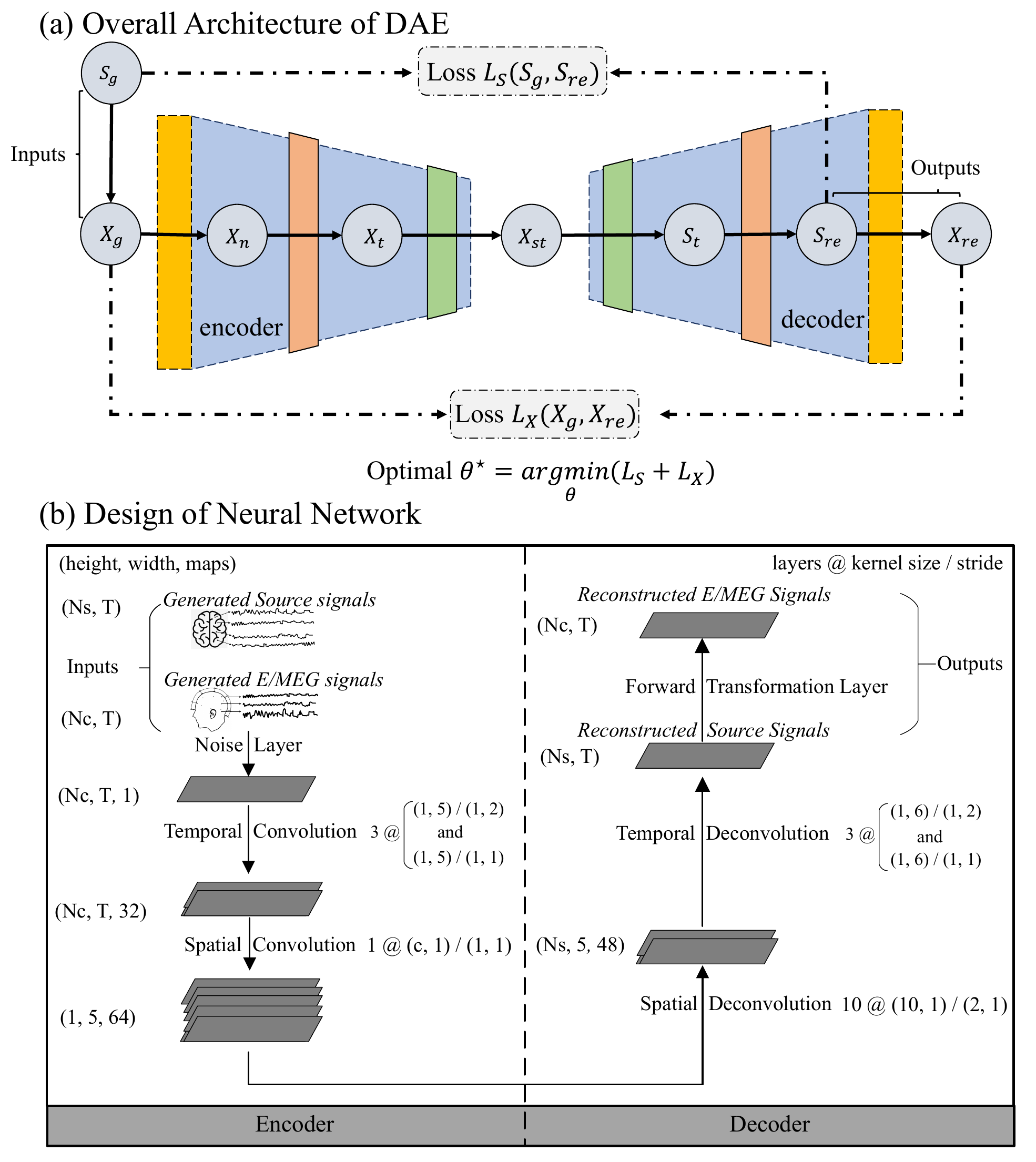}
		\caption{Overall architecture and design of the CedNet. (a) Overall architecture. The arrows indicate the input-to-output sequence of the CedNet, and the dashed lines indicate the connected tensors to calculate the corresponding loss function. (b) Detailed design of the neural network. $N_c$ and $N_s$ represent the number of sensors and sources, respectively. $T$ is the total number of time points. In our experiments, we downsampled the time points to 40.}
		%	\hfil
		\label{fig_2}
	\end{figure}

    % modified in 6/16
	The overall network structure is shown in Fig. \ref{fig_2}(a). The generated E/MEG signals $\bm{X}_g$ and the corresponding generated source signals $\bm{S}_g$ are fed into the neural network as training samples, and $\bm{X}_n$ is the noise corrupted version of $\bm{X}_g$. The encoder contains a noise corruption step and two feature extraction blocks: temporal and spatial encoding blocks. Specifically, through the temporal encoding block, the temporal feature $\bm{X}_t$ is extracted from the noise corrupted E/MEG signals afforded by the noise layer. 
	% modified in 6/27
	The spatial block then generates the spatio-temporal feature $\bm{X}_{st}$. 
	% Then, the spatio-temporal principal feature $\bm{X}_{st}$ is consequently obtained through the spatial encoding block. 	
	In the decoder, which consists of spatial and temporal decoding blocks and a forward transformation layer, the spatial decoding block first decodes spatial information from $\bm{X}_{st}$, yielding the source temporal features $\bm{S}_t$. The temporal block then computes the reconstructed source signal $\bm{S}_{re}$ by decoding the temporal information. Finally, the forward transformation layer calculates the reconstructed scalp signal $\bm{X}_{re}$ as the output of the CedNet.
	%$\bm{X}_{st}$ is firstly decoded temporal information, termed as the source temporal feature $\bm{S}_t$ through the spatial decoding block.
	% $\bm{S}_t$ is further temporally decoded for the reconstruction of the source $\bm{S}_{re}$ through the temporal block. 
	% Finally, as the output of the network, the reconstructed scalp signal $\bm{X}_{re}$ is calculated using the forward model $\bm{L}$. 
	\textbf{Unlike in the traditional CedNet, here the loss function is the weighted sum of the discrepancy between $\bm{S}_{re}$ and $\bm{S}_{g}$ and the discrepancy of $\bm{X}_{re}$ and $\bm{X}_{g}$. The former measures the error between the predicted and true response variables from the perspective of multi-target regression, hence falling under the umbrella of discriminative learning. 
    % update
	The latter introduces an auxiliary learning task which measures the adequacy of the CedNet for latent-space representation.
% 	The latter aligns with the loss function in the traditional DAE, which measures the adequacy of the DAE for latent-space representation. 
	}

\subsubsection{Encoder and Decoder}
	The details of the neural network are illustrated in Fig. \ref{fig_2}(b). The network is comprised of the following blocks:
	
	\textbf{Noise corruption} $\mathfrak{N} \ ( \mathbb{R}^{N_c \times T} \rightarrow \mathbb{R}^{N_c \times T})$: This step yields $\bm{X}_n$ by corrupting $\bm{X}_g$ with white Gaussian noise with a specific SNR. Note this block is only used during the training stage.
	
	% contains a generation process of white Gaussian noise and uses the generated noises to corrupt $\bm{X}_g$ into $\bm{X}_n$ with the specified SNR.
	
	% modified in 6/28
	\textbf{Temporal encoding block} $\mathfrak{h} \ ( \mathbb{R}^{N_c \times T} \rightarrow \mathbb{R}^{N_c \times K_t \times F_1} )$: This block contains multiple stacks of 2-D convolution layers with a convolution kernel of $1\times5$ and a step size of $1\times2$, where a convolution layer with the same kernel size but a $1\times1$ step size is added after each layer to enhance the modeling capability. The kernel size allows the block to perform temporal convolution separately for each channel of E/MEG signals, thus extracting the temporal features $\bm{X}_t=\mathfrak{h}(\bm{X}_n)$ that contains $F_1$ temporal feature maps with a temporal size of $K_t$. 

	\textbf{Spatial encoding block} $\mathfrak{g} \ (\mathbb{R}^{ N_c \times K_t \times F_1} \rightarrow \mathbb{R}^{ 1\times K_t \times K_s})$: This block contains a 2-D convolution layer with a convolution kernel of $N_c\times1$ and a step size of $1\times1$. Therefore, it performs a full-channel spatial convolution at each time point and extracts the spatio-temporal feature maps $\bm{X}_{st}=\mathfrak{g}(\bm{X}_t)$ with a size of $K_s \times K_t $.
	
	\textbf{Spatial decoding block} $\mathfrak{g}^{-1} \ (\mathbb{R}^{1\times K_t \times K_s } \rightarrow \mathbb{R}^{ N_s \times K_t \times F_2}) $: This block contains multiple stacks of 2-D deconvolution (also known as transposed convolution) layers with a kernel size of $10\times1$ and a step size of $2\times1$ applied to the feature maps along the spatial dimension. Each deconvolution layer performs an inverse convolution operation that swaps the forward and backward passes of a convolution, increasing the size of the spatial dimension of the feature maps\cite{dumoulin2016guide}. Collectively, the decovolution layers upsample the size of the spatial dimension from $1$ to $N_s$ with a 2-fold increase per layer. Therefore, this block decodes spatial information from $\bm{X}_{st}$ to yield the source temporal features $\bm{S}_t = \mathfrak{g}^{-1} (\bm{X}_{st})$ that contain $F_2$ feature maps with a size of $N_s\times K_t$.

	\textbf{Temporal decoding block} $\mathfrak{h}^{-1} \ (\mathbb{R}^{N_s \times K_t \times F_2} \rightarrow \mathbb{R}^{N_s \times T})$: This block contains multiple 2-D deconvolution layers with a kernel size of $1\times 6$ and a step size of $1\times2$ applied to the feature maps along the temporal dimension, upsampling its size from $K_t$ to $T$ with a 2-fold increase per layer. To enhance the capacity of the model, a deconvolution layer with the same kernel size but a $1\times1$ step size is also added after each layer. We set the kernel size of the deconvolution layer to an even number (i.e., 6) to avoid checkerboard artifacts\cite{odena2016deconvolution}. This block further decodes $\bm{S}_t$ to estimate the reconstructed source signals $\bm{S}_{re}=\mathfrak{h}^{-1} (\bm{S}_t)$.

%	This block contains multiple 2-D deconvolution layers with a kernel size of $1\times 6$ and a step size of $1\times2$, mapping the size of temporal axis from $K_t$ to original $T$ time points in a 2x (step size) fold and independently filter feature maps along the spatial axis, where a deconvolution layer with the same kernel size but a $1\times1$ step size is also added after each layer.
	
	%It implements the temporal transposed convolution at each source independently
	
	\textbf{Forward transformation block} $\mathfrak{f} \ (\mathbb{R}^{N_s \times T} \rightarrow \mathbb{R}^{N_c \times T})$: This block calculates the reconstructed E/MEG signals through the lead field matrix: $\bm{X}_{re}=\bm{L}\bm{S}_{re}$. 
	
	% The introduction of $\bm{X}_{re}$ in network training can efficiently improve the training performance. 

	To accelerate network training, the exponential linear unit ($\bm{elu}$) function\cite{clevert2015fast} is employed as the activation function for all convolution and deconvolution layers except the last deconvolution layer preceding $S_t$, which uses the linear activation function instead. 
	% modified in 6/28
	Furthermore, as suggested in \cite{ioffe2015batch}, a Batch Normalization (BN) layer is added between the convolution/deconvolution layer and the activation layer. This overcomes the distribution shift and the covariate drift that may occur during feature transformation especially in training with a large-scale dataset, preventing loss oscillation, weight saturation, and decelerated learning. 
	According to our experiments, with the BN layer introduced in the neural network, the solution is not only insensitive to the parameter initialization in each layer but also able to converge quickly and smoothly.
	
	% modified in 6/28: delete the explaination of why we add BN layers
	% due to the tremendous diversity among the generated samples, a large-scale-training strategy for parameter initialization is difficult to implement. In addition, because of the oscillating training loss induced by data diversity, the network cannot easily convergence. Therefore,
	% which is desirable since shift and drift lead to

	\subsubsection{Loss Function}
	
	In this paper, two error functions quantifying the reconstruction error for the source signals and that of the E/MEG signals are proposed.
	%the data-discrepancy function of measurements $D(\bm{X}_g, \bm{S}_g)$ and the auxiliary regularization-like function of source $\bm{\bar}{R}(\bm{S}_g)$, are proposed, 
	The  loss function is the weighted sum of the two error functions: 
	\begin{equation}
	\begin{split}
	\mathcal{L}_{\theta}(X_g,S_g)
%	& = \lambda_1 D(\bm{X}_g, \bm{S}_g) + \lambda_2 \bm{\bar}{R}(\bm{S}_g)\\
	& = \lambda_1 \mathcal{L}_X(\bm{X}_g,\bm{X}_{re}) + \lambda_2 \mathcal{L}_{S}(\bm{S}_g,\bm{S}_{re}), \\
	%				& = \lambda_1 L_X + \lambda_2 (L_{S_1} + \delta L_{S_2})
	\end{split}
	\end{equation}
	where $\lambda_1 \ge 0$ and $\lambda_2 \ge 0$ are the loss weights (see Supplementary Materials for further discussions), and 
	
	% \begin{equation}
	% \mathcal{L}_X(\bm{X}_g,\bm{X}_{re})=\frac{1}{N} \frac{1}{N_c} \sum_{i}^{N} \sum_{q}^{N_c} \|\bm{X}_{g}^{(i)}-\bm{X}_{re}^{(i)}\|_\mathcal{F} ^2,
	% \end{equation}
	
	% \begin{equation}
	% 	\mathcal{L}_X(\bm{X}_g,\bm{X}_{re})=\frac{1}{N} \frac{1}{T} \sum_{i}^{N} \sum_{t}^{T} \| \bm{X}_{g}^{(i)}(t)-\bm{X}_{re}^{(i)}(t) \|^2_2,
	% \end{equation}

	\begin{equation}
		\mathcal{L}_X(\bm{X}_g,\bm{X}_{re})=\frac{1}{N} \sum_{i}^{N} \| \bm{X}_{g}^{(i)}-\bm{X}_{re}^{(i)} \|_\mathsf{F} ^2,
	\end{equation}
	%_\mathcal{F} 

	%  matrix norm 
	\begin{equation}
	\mathcal{L}_{S}(\bm{S}_g,\bm{S}_{re})=\frac{1}{N} \sum_{i}^{N}
	\left[ \|\bm{S}_{g}^{(i)}-\bm{S}_{re}^{(i)}\|_\mathsf{F} ^2 + \delta \| \bm{S}_{g}^{(i)}-\bm{S}_{re}^{(i)} \|_{1,1} \right],
	\end{equation}
	% vector norm
	% \begin{equation}
	% 	\begin{aligned}
	% 	\mathcal{L}_{S}(\bm{S}_g,\bm{S}_{re})=\frac{1}{N} \frac{1}{T} \sum_{i}^{N} \sum_{t}^{T}\left[ { \| \bm{S}_{g}^{(i)}(t)-\bm{S}_{re}^{(i)}(t)\|^2_2 } \right. \\
	% 	\left. { +\delta \| \bm{S}_{g}^{(i)}(t)-\bm{S}_{re}^{(i)}(t) \|_1 } \right],
	% 	\end{aligned}
	% \end{equation}
	% + \delta \| \bm{S}_{g}^{(i)}(t)-\bm{S}_{re}^{(i)}(t) \|_1
	%_\mathcal{F}  _{1,1}
	%$X(t) = \bm{x}_t$, $S(t) =\bm{s}_t$, 
	where $\| \cdot \|_{1,1}$ is the $L_{1,1}$ matrix norm defined as $\| \bm{A} \|_{1,1} = \sum_{j = 1}^{n}\sum_{i = 1}^{m}|a_{ij}|$; $N$ is the total number of training samples and $\delta \ge 0$ is a manually set parameter. To facilitate precise reconstruction of the sources by the network, we use the weighted sum of the mean square error (MSE) and mean absolute error (MAE) to assess the reconstruction error for the source signals, since the MSE is more sensitive to high amplitude errors while the MAE is more sensitive to subtle differences in the amplitude. However, for the E/MEG signals, due to the presence of noise in the data, excessively pursuing the reconstruction accuracy may lead to overfitting. Therefore, we use the mean square error (MSE) alone to measure the reconstruction error.  
    %update	
	\textbf{In contrast to the standard CedNet that only attempts to learn the source signals as in Equation (6), which are what the ESI strives to estimate, the loss function for DST-CedNet additionally incorporates the optimization of latent-space representation for E/MEG (Equation (5)), which further enhance the denoising capability for the model.} 
% In contrast to the standard DAE that only attempts to learn a latent-space representation for E/MEG (Equation (5)), the loss function for DST-DAE additionally incorporates the optimization of the source signals as in Equation (6), which are what the ESI strives to estimate.
	Moreover, since the generated source signals encode our prior knowledge regrading the true dynamics of brain activities, our approach provides a convenient framework for leveraging prior information regarding source signals as long as it is properly reflected in the data synthesis step. Thus, a notable difference of our approach from traditional ESI methods is that the prior information regarding source signals is not enforced via mathematical constraints typically presented as the $L_1$ norm or $L_2$ norm of the source signals for mathematical convenience, which nonetheless may not accurately reflect that the actual source configurations.

	Performance comparisons of different ESI algorithms are presented in Section \cref{sec.experiments.SS,sec.experiments.SSE,sec.experiments.SP,sec.experiments.CC} to demonstrate the advantages of DST-CedNet.
	%It can effectively eliminate the approximation error induced by explicit prior modeling.
	
	\subsection{Implementation Details} \label{sec.methodology.IP}
	The workflow of DST-CedNet consists of three phases: data synthesis phase, training phase and source estimation phase. 
	%The pseudo-code of the algorithm is shown in the Appendix \ref{sec.AI.2}.
	\subsubsection{Data Synthesis Phase}
	To ensure that the training data faithfully reflect the spatio-temporal properties of realistic brain activity, the proposed data strategy incorporate spatial and temporal information of brain activity into the generation of source signals:
		\begin{equation}
	\bm{S}_g =\bm{W} \bm{\Phi} = \sum_{k=1}^{K} \bm{w}_k \bm{\phi}_k ,
	\end{equation}
	where $\bm{\Phi}  \in \mathbb{R}^{K\times T}$ is the temporal state matrix and $\bm{W}\in \mathbb{R}^{N_s\times K}$ represents the activation state matrix of brain sources with $K$ spatial components. $\bm{\phi_k} $ is the $k$-th temporal component generated from a set of $m$ temporal basis functions (TBFs) $\bm{\Theta}\in\mathbb{R}^{m\times T}$. $\bm{w}_k$ is the $k$-th spatial component:

	%= [\phi_1,\cdots,\phi_j,\cdots,\phi_K]^\mathsf{T}
	%= [w1,\cdots,w_j,\cdots,w_K]
	\begin{equation}
	\bm{w}_k=
	\begin{cases}
	1& s_j\in \bm{\gamma}\\
	0& \text{else}
	\end{cases}
	,
	\end{equation}
	where $s_j$ represents the $j$-th vertex of the source, and $\bm{\gamma}$ represents the index set of activated sources.
	
	For the purpose of large-scale training, we generate training samples with varying spatial activation states $\bm{W}$ and temporal states $\bm{\Phi}$ as follows. For the $i$-th generated sample, first, we sample from a uniform distribution to obtain the seed source $s_{seed}$ in the source space. Subsequently, the spatial activation state is generated as a binary vector whose activated sources are indexed as 1, wherein the index set of activated sources is obtained by gradually expanding activated sources within the seed's neighborhood $n$ until the total area is subject to a given value $A$: $\bm{\gamma}^{(i)}_k=\{s_j|area(\bm{\gamma})= A ,s_j \in n(s_{seed}) \}$, where $A$ is randomly sampled from a uniform distribution between 0 and $A_{max}$. $A_{max}$ is set to $20$ as it has been shown the area of cortex activity is typically $10–20$ $cm^2$\cite{tao2005}. The generated spatial states can be spatially contiguous and locally homogeneous as suggested by previous studies\cite{hamalainen1993,destexhe1999}.
	
	Second, methods such as the stimulus evoked factor analysis (SEFA)\cite{nagarajan2007probabilistic} or singular value decomposition (SVD)\cite{ou2009distributed} can be used to estimate the TBFs from realistic E/MEG signals $X_{real}$ (see Sections \ref{sec.experiments.SE} and \ref{sec.experiments.MEG}). Then, $\bm{\phi_k} ^{(i)}$ is generated from the subspace spanned by $\bm{\Theta}$ as $\bm{\phi_k} ^{(i)}= \sum_{m=1}^{M} \zeta_m ^{(i)} \bm{\Theta}(m) $, where the weight $ \zeta_m ^{(i)} $ is obtained from uniform sampling.

	Finally, the $i$-th generated source signals $\bm{S}_g^{(i)}$ can be obtained as the product of the spatial activation state $\bm{W}^{(i)}$ and the temporal state $\bm{\Phi}^{(i)}$, and the corresponding E/MEG signals $ \bm{X}_g^{(i)}$ can be generated by passing the generated source signals through the forward model with the lead field matrix $\bm{L}$ derived from solving the forward problem.
	
	\subsubsection{Training Phase}
	The generated dataset ${(\bm{S}_g, \bm{X}_g)}$ is used to train the neural network, where the SNR of the E/MEG signals determines the noise level in the noise corruption to corrupt $X_{g}^{(i)}$. Through minimizing the loss function until convergence, the trained neural network is capable of automatically learning the mapping from the E/MEG signals to the source signals.

	\subsubsection{Source Estimation Phase} 
	In this phase, the trained neural network is applied to map unseen realistic E/MEG signals $\bm{X}_{real}$ to source signals $\bm{S}^ \star$.

	% exp
	\section{Experiments}
	\label{sec.experiments}
	In this section, we present experiments on both simulated and realistic data to demonstrate the efficacy of DST-CedNet. Specifically, four simulated experiments were conducted to compare the performance of DST-CedNet to that of several existing ESI methods, including two $L_2$-norm methods, i.e., wMNE\cite{Dale1993} and LORETA\cite{PascualMarqui1994}, which were proposed for the reconstruction of spatially extended source, three sparse-constrained methods, i.e., SBL\cite{Wipf2010}, FOCUSS\cite{Gorodnitsky1995}, and Champagne \cite{Wipf2010,cai2020robust}, which were devised for estimating focal sources, and spatio-temporally constrained method, STTONNICA\cite{ValdsSosa2009}, which decomposes source signals into spatial and temporal components that are optimized under the Bayesian information criterion (BIC). \textbf{Additionally, two recent deep learning algorithms are also 
   included for comparison, namely SIFNet \cite{Sun2020sifnet} and ESBN \cite{wei2021edge}. Specifically, SIFNet reformulates the ESI problem as a multi-class classification problem. Although temporal information is utilized for data synthesis, SIFNet does not reconstruct the temporal dynamics of the sources but rather focuses on the localization of sources. ESBN treats source activities as independently and identically distributed. Since SIFNet is only able to determine the activation states of the sources as opposed to their temporal dynamics, we calculated all of the performance metrics (i.e., AUC, SD, and DLE) based on the peak values of the ground-truth source activities and those of the reconstructed source activities. As a result of hyperparameter optimization on the training set, the convolutional kernel size of SIFNet is $3\times3$, and the learning rate is $10^{-5}$; For ESBN, the learning rate is $10^{-6}$,  and the loss weights are $1$, $10^{-3}$, $10^{-6}$, and $10^{-6}$ for the mean square error, cosine similarity, $L_{1}$ norm, and edge sparse norm, respectively. The other settings are identical to those used in \cite{Sun2020sifnet} for SIFNET and \cite{wei2021edge} for ESBN.} The performance of DST-CedNet is assessed by evaluating the recovered source signals under a variety of experimental settings using multiple metrics. An additional experiment was conducted to study the impact of training sample size on the performance of DAT-CedNet. \textbf{Finally, real MEG and EEG datasets were analyzed to further illustrate the efficacy of DST-CedNet on real-world data.}
% 	Finally, a real MEG dataset was analyzed to further illustrate the superiority of DST-CedNet in practice.

	\subsection{Experimental Description and Performance Metrics } \label{sec.methodology.ED}
	
	\subsubsection{Protocol of Simulation Experiments}
	\label{sec.experiments.SE}

	Throughout the following experiments, the cortical surface was obtained by segmenting the white/gray matter interface of default MR images derived from MNI/ICBM152 using Brainvisa\footnote{http://www.brainvisa.info.}. The cortical surface was then tessellated with triangles and the resulting cortical mesh was downsampled to a total of 1,024 triangular elements in order to build the cortex space. The sensor template was configured as the 65-channel Neuroscan Quikcap system. The lead field matrix $\bm{L}$ was computed using the BEM method. This forward modeling process was implemented using the OpenMEEG plug-in\cite{gramfort2010openmeeg} in Brainstorm \cite{tadel2011brainstorm}. 
	
	The time course of each patch source was simulated as a Gaussian damped sinusoidal time course: $sin(2\pi f) exp^{-(t-{\tau}/{\omega})^2}$. Through the choice of different $f$, $\omega$, and $\tau$, event-related potential (ERP) signals with various temporal states could be simulated. A source active region was simulated by selecting a triangular element on the cortical mesh as a seed voxel in the regional center and enlarging the region by iteratively absorbing its neighbor elements until the required area condition was achieved. The source signals were constructed by placing the simulated ERP signals at the source active regions, and the corresponding EEG signals were calculated via the forward model. White Gaussian noise with a specific SNR was added to the simulated source and scalp signals. The SNR was defined as $SNR=10 \log_{10} \frac{\|\bm{LS}\|^2}{\|\bm{\varepsilon}\|^2}$. In our simulation experiments, we generated the ERP signal with 40 sampling points per trial, and the TBFs were obtained from the simulated E/MEG signals by SEFA. Without loss of generality, the number of TBFs was 4.
	
	In simulation experiments, we considered four scenarios to assess the reconstruction performance of DST-CedNet against aforementioned compared methods:
	
	\begin{itemize}
		
		\item Single active patch source with varying SNRs: we fixed the source area at 5$\pm$0.58 ${cm}^2$ and varied the SNR from -5 dB, 0 dB, 5 dB, to 10 dB.
		\item Single active patch source with varying spatial extents: we fixed the SNR at -5 dB and varied the source area from 3${cm}^2$, 6 ${cm}^2$, 10 ${cm}^2$, to 15 ${cm}^2$.
		\item Two patch sources with varying spatial patterns: we fixed the SNR at -5 dB and varied the source pattern from 4 ${cm}^2$ and 4 ${cm}^2$, 4 ${cm}^2$ and 10 ${cm}^2$, 10 ${cm}^2$ and 4 ${cm}^2$, to 10 ${cm}^2$ and 10 ${cm}^2$.
		\item Two patch sources with varying correlated sources: we fixed SNR at -5 dB and the source area at 7 ${cm}^2$. The coefficient correlation between the two patch source signals was varied from 0, 0.3, 0.6, to 0.9.
		
	\end{itemize}
	
	In each experiment, the training set and validation set were generated with an equal sample size of $N = 27,000$ using our data synthesis approach. Each trial of 64-channel synthetic signals contained 40 time points and were corrupted by a randomly generated additive noise with an SNR of -5 dB during the training of the convolutional encoder-decoder network. The source space consisted of 1,024 vertices.  Subsequently,  the model was optimized on the training set with settings described in the next subsection. The predictive performance of the trained models was then assessed on the validation set. A total of 100 Monte Carlo simulations were performed to statistically evaluate the reconstruction performance.

	\subsubsection{Parameter Settings}
	The proposed network was coded in Python with Keras library using TensorFlow Backend. Adam, which is a gradient-based optimization algorithm, was chosen as the optimizer. The impact of the hyperparameters on the predictive performance of the neural network was systematically studied on separately generated data (see Section \ref{sec.discussion.HT} for details). In short, the predictive performance of the neural network favors the following settings of the key hyperparameters: 1) small kernel sizes and small strides in the temporal encoding and decoding layers; 2) large kernel sizes and small strides in the spatial encoding and decoding layers. In particular, a kernel size equal to the channel number in the spatial convolution layer is preferred; and 3) increasing numbers of feature maps for the encoder and decreasing numbers of feature maps for the decoder. These observations therefore motivated us to set the kernel size and stride as in Fig. \ref{fig_2}. Moreover, we set $K_t=5$ and $K_s=64$ for the spatio-temporal feature $\bm{X}_{st}$. The optimal ratio of loss weights are $\lambda_2:\lambda_1 = 15:1$. The learning rate of the Adam optimizer was set to $10^{-4}$, and the other hyperparameters of the Adam optimizer were set to the values suggested in\cite{kingma2014adam}, i.e., $\beta_1=0.9$, $\beta_2=0.999$ and $\epsilon=10^{-8}$. The batch size was set to 32 for the mini-batch stochastic gradient descent, and all the trainable parameters in the neural network were initialized with the Glorot Uniform method. The epoch number of the model training was set to 250. We used a weight decay strategy for every convolution/deconvolution layer by employing the $L_1$ and $L_2$ regularizers, for which the regularization penalty is $10^{-3}$. $\delta$ in the loss function was set to 0.1.  
	
	To further improve the performance of DST-CedNet, we utilized a warm-up strategy\cite{gotmare2018closer} to dynamically fine-tune the learning rate during the training. The learning rate was linearly increased to the maximum in the initial 20 epochs and linearly decreased to the minimum in the last 20 epochs.

	\subsubsection{Performance Metrics}
	In the simulation experiments, the performance of all the ESI methods was assessed with the following four metrics:
	\begin{itemize}
		\item Area under the receiver operating characteristic curve (AUC), which evaluates the sensitivity and specificity of the source detection.
		
		\item Relative mean square error (RMSE), which evaluates the accuracy of the reconstructed time courses.
		
		\item Distance of localization error (DLE), which measures the localization error through the average distance from the activity peak of the estimated sources to those of the true ones.
		
		\item Spatial dispersion (SD), which measures the extent to which the estimated sources are spatially blurred.
	\end{itemize}
These metrics allow us to comprehensively evaluate each ESI method in terms of its ability to localize and estimate the spatial extent of the sources. In general, a desirable method is expected to achieve a relatively high AUC value and small RMSE, SD, and DLE values. Furthermore, as an indicator of the generalization performance of the trained network, we calculated the R-square ($R^2$) \cite{alexander2015beware} values between the estimated and true source signals in the validation set unused in the training stage. Source maps were thresholded using the Otsu's method \cite{otsu1979threshold}. Details of each metric and the Otsu's method are provided in Supplementary Materials.
	
%	 \cite{cameron1997r} r square

	\subsection{Experiment I: Single Patch Source with Varying SNRs} \label{sec.experiments.SS}
	\begin{figure}[htbp]
		\centering
		\includegraphics[width=3.4in]{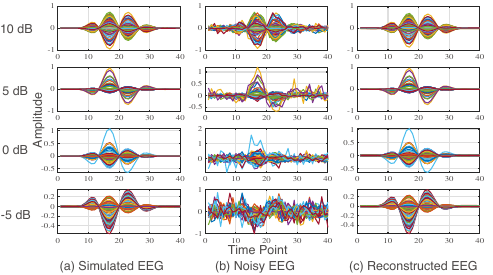}
		\caption{Denoising performance of CedNet for EEG signals with varying SNRs. The left column represents noiseless EEG signals randomly sampled from each group of test samples. The middle column represents the corresponding noisy signals. The right column represents the reconstructed signals that are denoised via the neural network.}
		%	\hfil
		\label{fig_3}
	\end{figure}

	\begin{table}
        \setlength\tabcolsep{4pt} %left 3pt
		\centering
     	\caption{RMSE values between simulated EEG signals and reconstructed signals. The mean and standard deviation are calculated by 100 Monte Carlo simulations per SNR.}
		\begin{tabular}{lcccc}
			\toprule
			SNR (dB) & -5 & 0 & 5 & 10 \\
			\midrule
			RMSE & 0.781$\pm$0.036 & 0.519$\pm$0.029 & 0.269$\pm$0.015 & 0.132$\pm$0.005 \\
			\bottomrule
		\end{tabular}

		\label{Tab_2}
	\end{table}

\begin{figure}[h]
		\centering
		\includegraphics[width=0.475\textwidth]{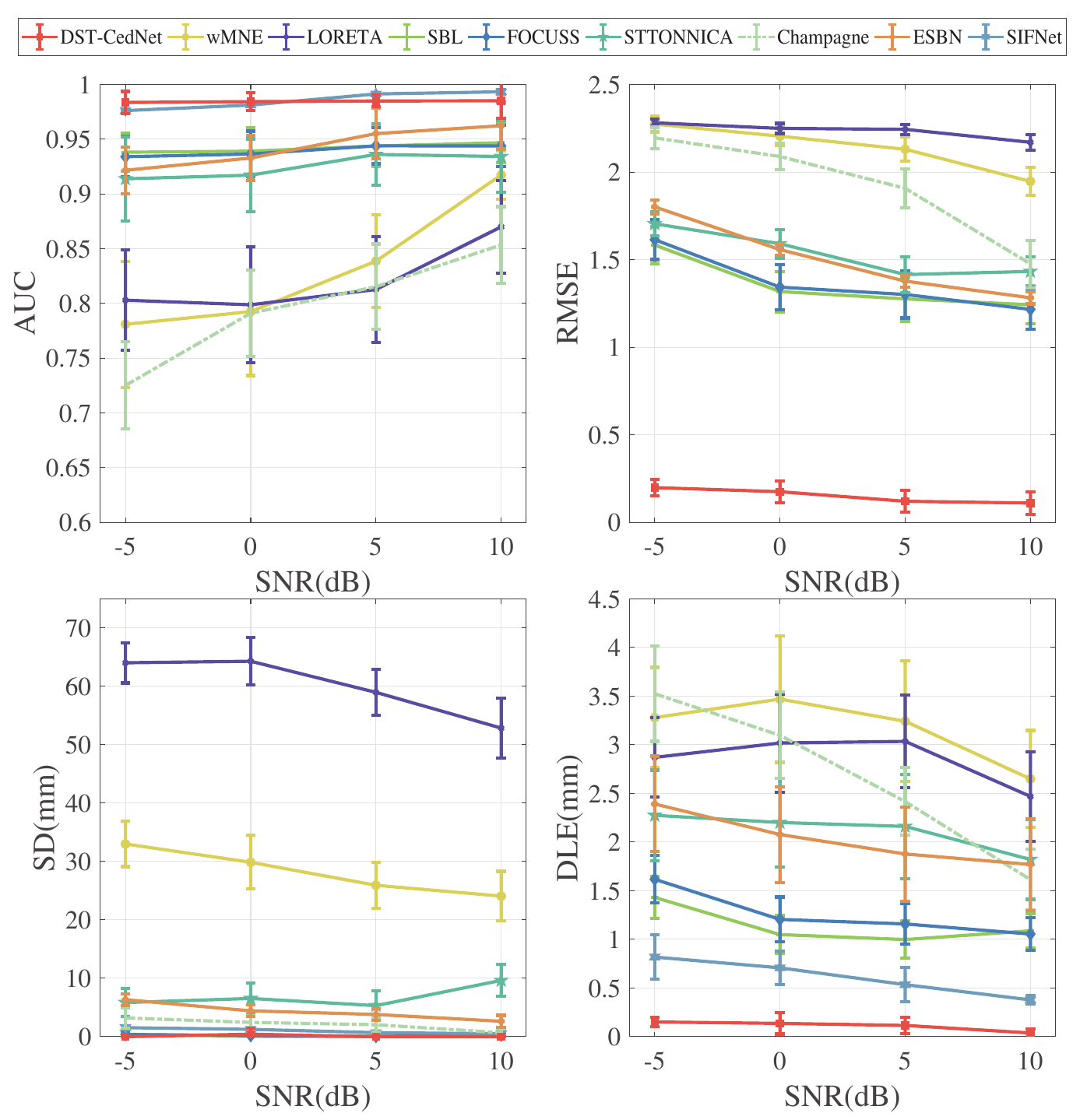}
		\caption{Performance metrics with varying SNRs. Results are assessed across 100 Monte Carlo simulations and shown as mean $\pm$ SEM (SEM: standard error of the mean). There are four metrics: 1) AUC evaluates the sensitivity and specificity of source localization, 2) RMSE evaluates the accuracy of the reconstructed time courses, 3) SD measures the spatial blurred extent, and 4) the DLE quantifies the localization error.}
		%		\hfil
		\label{fig_4}
	\end{figure}

	The first experiment evaluates the reconstruction performance for source signals and EEG signals with varying SNRs when the spatial extent of the patch source is fixed at 5 $cm^2$. As shown in Fig. \ref{fig_3} and Table \ref{Tab_2}, DST-CedNet can accurately reconstruct the time courses of the EEG signals even with heavy noise in the data. 
	Fig. \ref{fig_4} depicts the four performance metrics over 100 Monte Carlo simulations. For the AUC value, the performance of wMNE, LORETA, ESBN, and Champagne worsens as the noise level increases, while STTONNICA, SBL, FOCUSS, and SIFNet are more robust to noise. 
	%update
	\textbf{As can be seen, although SIFNet achieves the highest AUCs when SNR is 5 dB or 10 dB, which is perhaps not surprising as it is designed to specifically solve ESI as a classification problem, Among all the methods, DST-CedNet achieves the highest AUC value under low SNRs ($\leq$ 0 dB), which remains stable across all the noise levels.}
	For the RMSE value, all the algorithms' performance declines with increasing noise. However, DST-CedNet substantially outperforms the other algorithms under all the SNRs with at least 5-fold improvement. Furthermore, DST-CedNet achieves low DLE and SD values as SBL and FOCUSS do. However, SBL and FOCUSS are sparse algorithms that favor focal sources at the price of higher RMSE and lower AUC values.

\begin{figure}[htbp]
		\centering
		\includegraphics[width=3.5in]{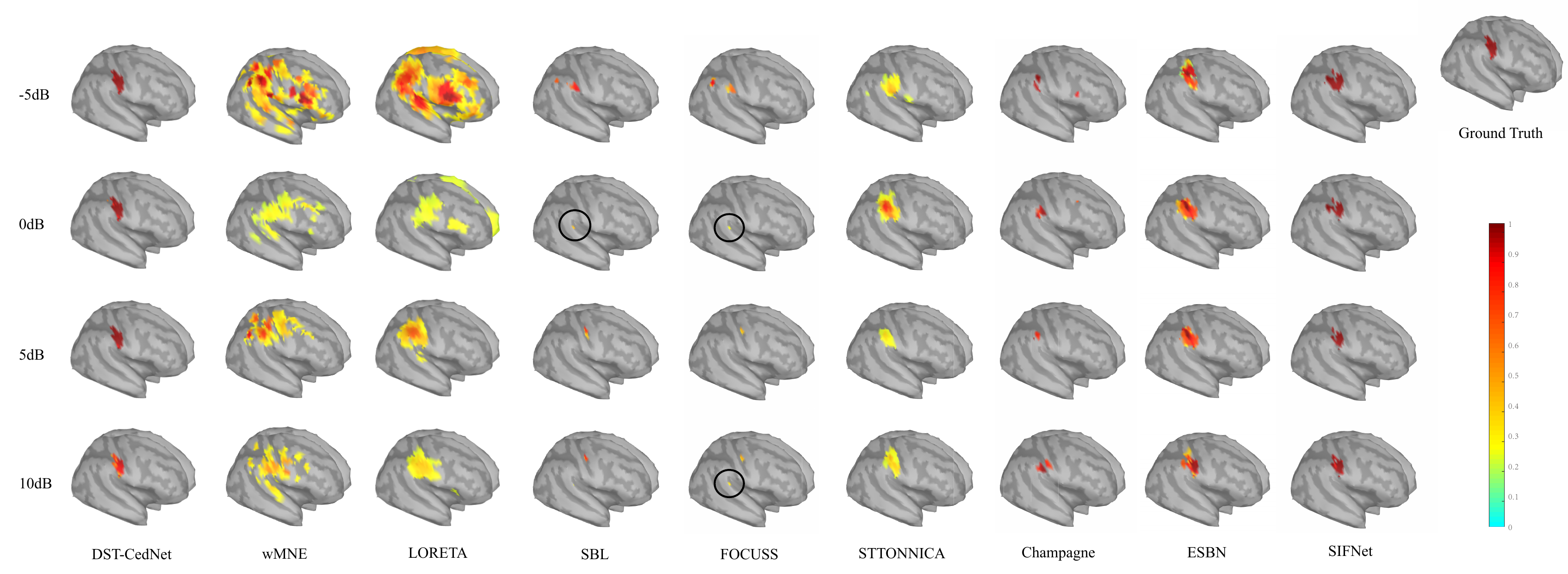}
		\caption{Source estimates with varying SNRs. Results are presented as the absolute values of the current activities at the peak of the simulated sources and thresholded using Otsu’s method. Areas of the sources are all 5 $cm^2$.}
		%	\hfil
		\label{fig_5}
	\end{figure}

	To illustrate, a source imaging example is presented in Fig. \ref{fig_5}. As can be seen, the $L_2$-norm based methods, namely wMNE and LORETA, lead to diffuse source estimates, which aggravates as the noise increases. By contrast, the sparse methods, namely SBL, FOCUSS, and Champagne, tend to split patch sources into focal sources, resulting in spurious sources with increasing noise. STTONNIC yields relatively compact sources. However, the amplitude of the sources is not accurate compared to the ground truth. 
	%update
	\textbf{Although ESBN and SIFNet can both accurately localize sources, errors are observable for the extent of the reconstructed sources with increasing noise.} 
	Since DST-CedNet does not make biased assumptions regarding the sources, it can faithfully recover the spatial extent of the sources with accurate estimation of their amplitude. Therefore, both the quantitative and qualitative results demonstrate the robustness of DST-CedNet compared to the other algorithms.

	\subsection{Experiment II: Single Patch Source with Varying Spatial Extents } \label{sec.experiments.SSE}
\begin{figure}[htbp]
		\centering
		\includegraphics[width=0.47\textwidth]{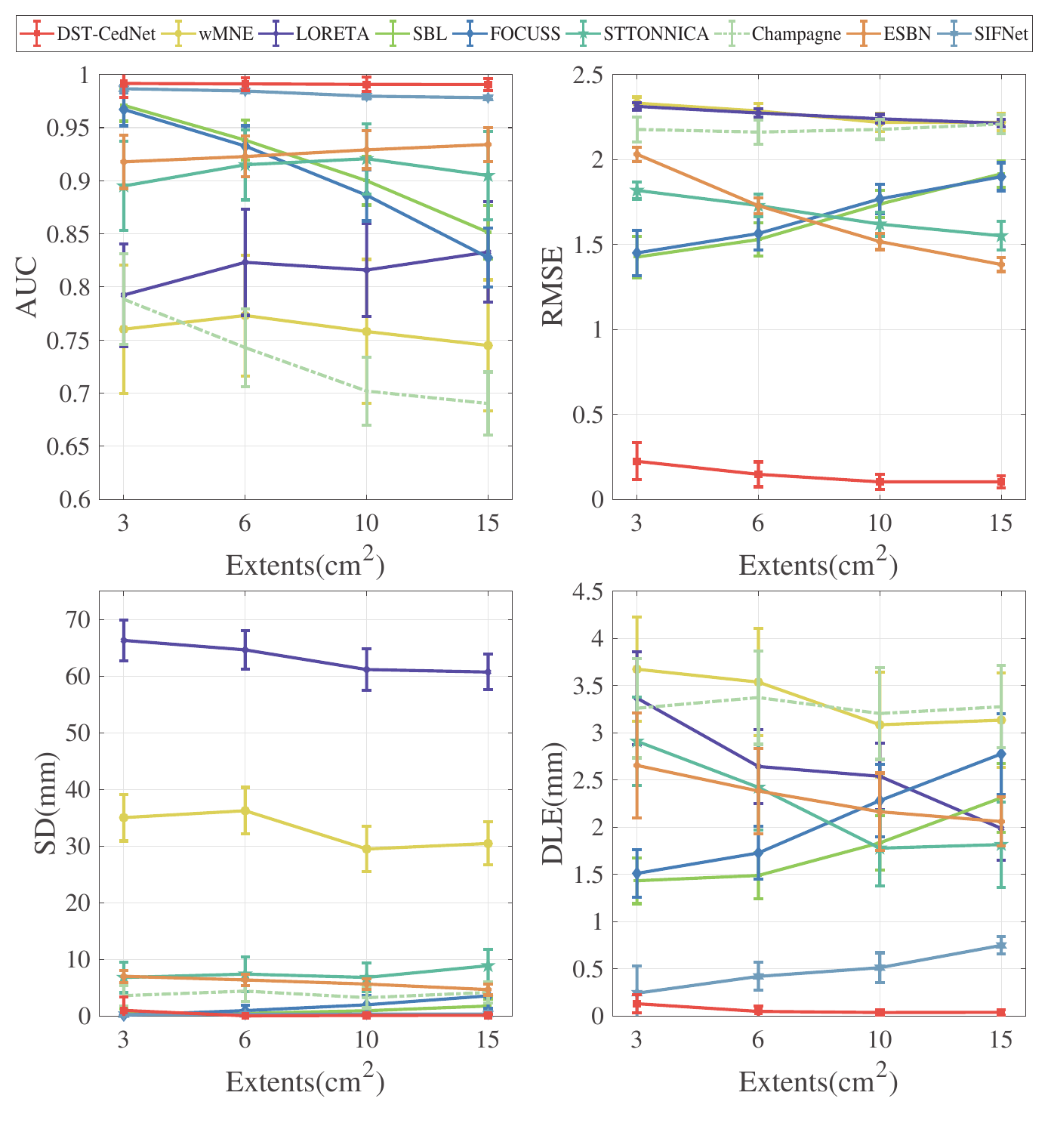}
		\caption{Performance metrics with varying spatial extents. Results are assessed across 100 Monte Carlo simulations and shown as mean $\pm$ SEM. SNR is -5 dB.}
		%	\hfil
		\label{fig_7}
	\end{figure}

    We next assess the performance of DST-CedNet for a single patch source with varying spatial extents when the SNR is fixed at -5 dB, which is arguably a challenging scenario for any ESI algorithms to handle. The performance metrics for the compared algorithms are shown in Fig \ref{fig_7}. Again, DST-CedNet achieves the highest AUC value and the lowest RMSE and DLE values across all the spatial extents. Moreover, despite the low SD value for FOCUSS and SBL when the spatial extent is less than 6 $cm^2$, the performance degrades when the spatial extent increases. However, DST-CedNet's SD remains low across all the spatial extents, demonstrating its ability to localize sources precisely with small spatial diffusion. 
    	
\begin{figure}[htbp]
		\centering
		\includegraphics[width=3.5in]{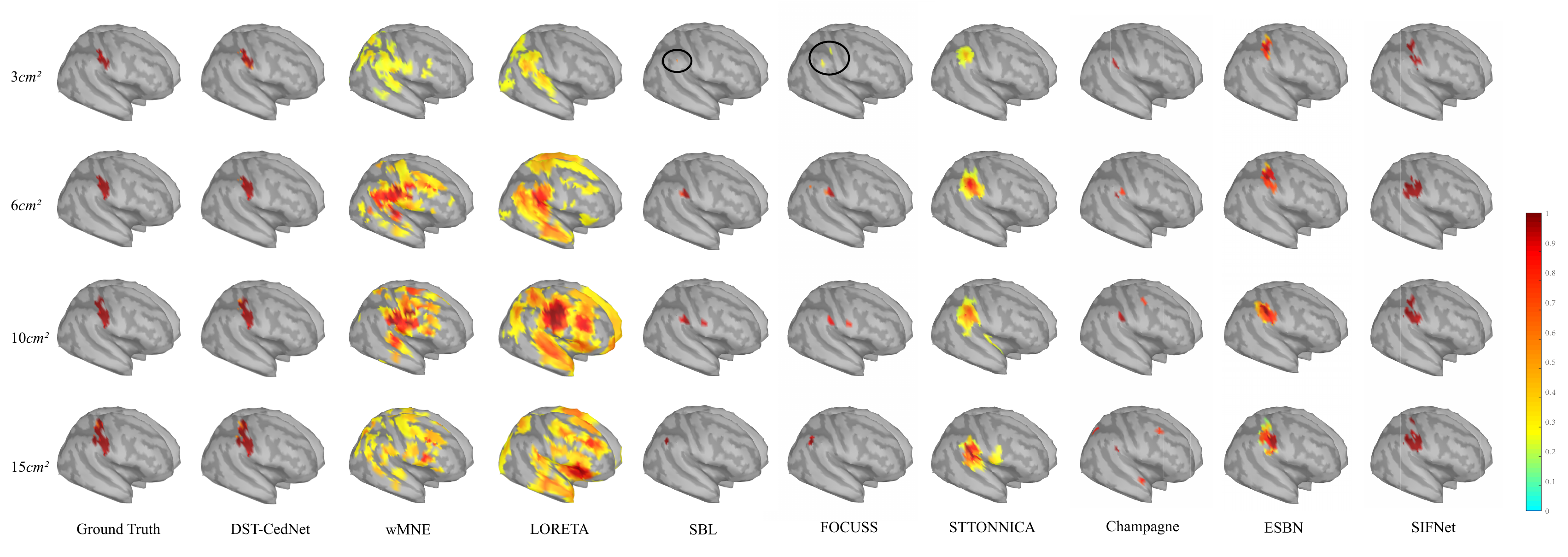}
		\caption{Source estimates with varying spatial extents. Results are presented as the absolute value of the current activities at the peak and thresholded using Otsu’s method.}
		%	\hfil
		\label{fig_6}
	\end{figure}

    Fig. \ref{fig_6} provides a source imaging example. The observations align with those in Fig. \ref{fig_5} in that SBL, FOCUSS, Champagne, wMNE, LORETA, and STTONNICA are unable to accurately estimate the spatial extent due to their biased prior assumptions;
    %update
    \textbf{
    Moreover, while the performance of ESBN enhances as the spatial extent of the sources increases, the estimates by SIFNet become fragmented.} By contrast, owing to the diversity of the training data comprising comprehensive spatio-temporal configurations, DST-CedNet can recover sources with accurate estimation of their locations and spatial extent.
    
    % Additionally, the performance of ESBN enhances as the spatial extent gets larger but SIFNET gets discontinuous source localization as such condition. However, owing to the diversity of the training data associated with abundant spatio-temporal information, DST-CedNet can recover sources with accurate estimation of their locations and spatial extent.

	%

	\subsection{Experiment III: Two Patch Sources with Varying Spatial Patterns} \label{sec.experiments.SP}
	\begin{figure}[htbp]
		\centering
		\includegraphics[width=0.47\textwidth]{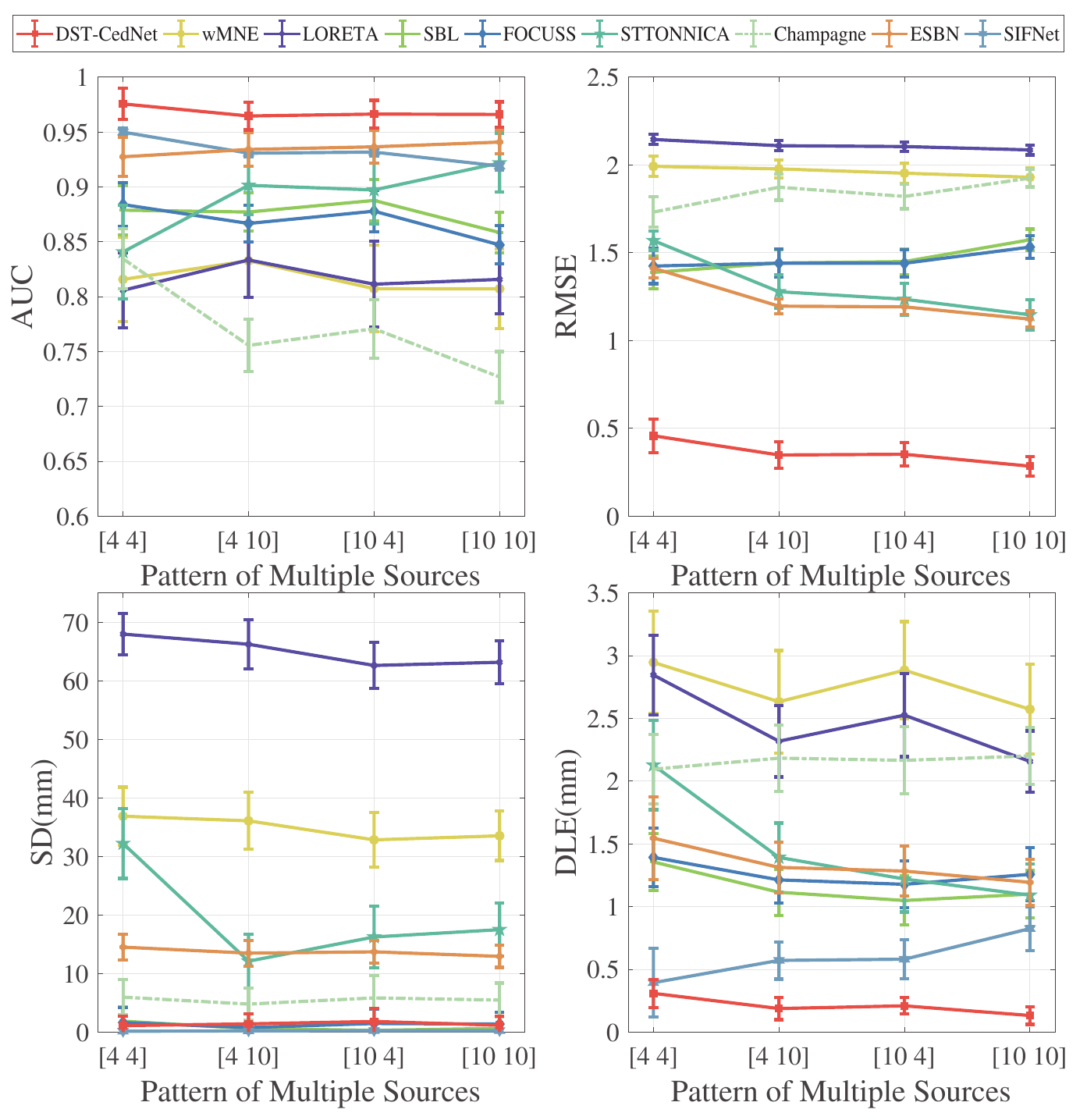}
		\caption{Performance metrics for two patch sources with varying spatial patterns. The x-axis represents different patterns of the two patch sources, where the numbers indicate the areas of the sources. Results are assessed across 100 Monte Carlo simulations and shown as mean $\pm$ SEM. SNR is -5 dB.}
		%	\hfil
		\label{fig_9}
	\end{figure}	
	\begin{figure}[htbp]
		\centering
		\includegraphics[width=3.5in]{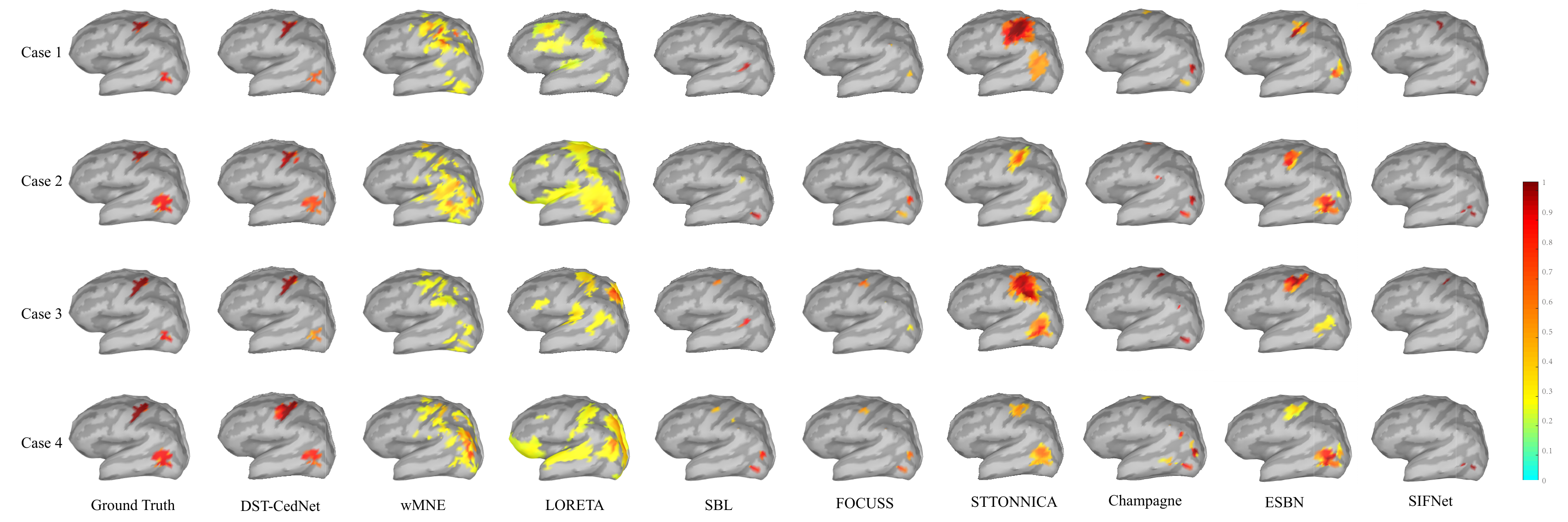}
		\caption{Source estimates for two patch sources with varying spatial patterns. Results are presented as the absolute value of the current density at the peak and thresholded using Otsu’s method.}
		%	\hfil
		\label{fig_8}
	\end{figure}
	
	In the third experiment we further evaluate the performance of different algorithms in the case of multiple patch sources. To simplify settings, we consider two patch sources, each with a different ERP time source that is independently generated. 
	% modified in 7/2
	In line with the results for the single patch source, DST-CedNet yields the best values across AUC, RMSE, and DLE. FOCUSS and SBL achieve slightly lower SD values than DST-CedNet for certain source areas (Fig. \ref{fig_9}.). However, these two algorithms favor focal sources at the price of sacrificing accuracy for estimating the spatial extents of the patch sources as reflected by their high DLE and RMSE values.

	As illustrated in Fig. \ref{fig_8}, the localization accuracy of all the algorithms declines when there are multiple patch sources. \textbf{Notably, as the extent of the sources gets larger, SIFNet misses a large portion of the sources. ESBN can localize multiple sources but their amplitudes are off compared to the ground truth. Among all the compared algorithms, DST-CedNet provides the best source recovery performance, demonstrating its capability of accurately reconstructing multiple patch sources.}

	\subsection{Experiment IV: Two Patch Sources with Varying Correlations } \label{sec.experiments.CC}
	\begin{figure}[htbp]
		\centering
		\includegraphics[width=0.47\textwidth]{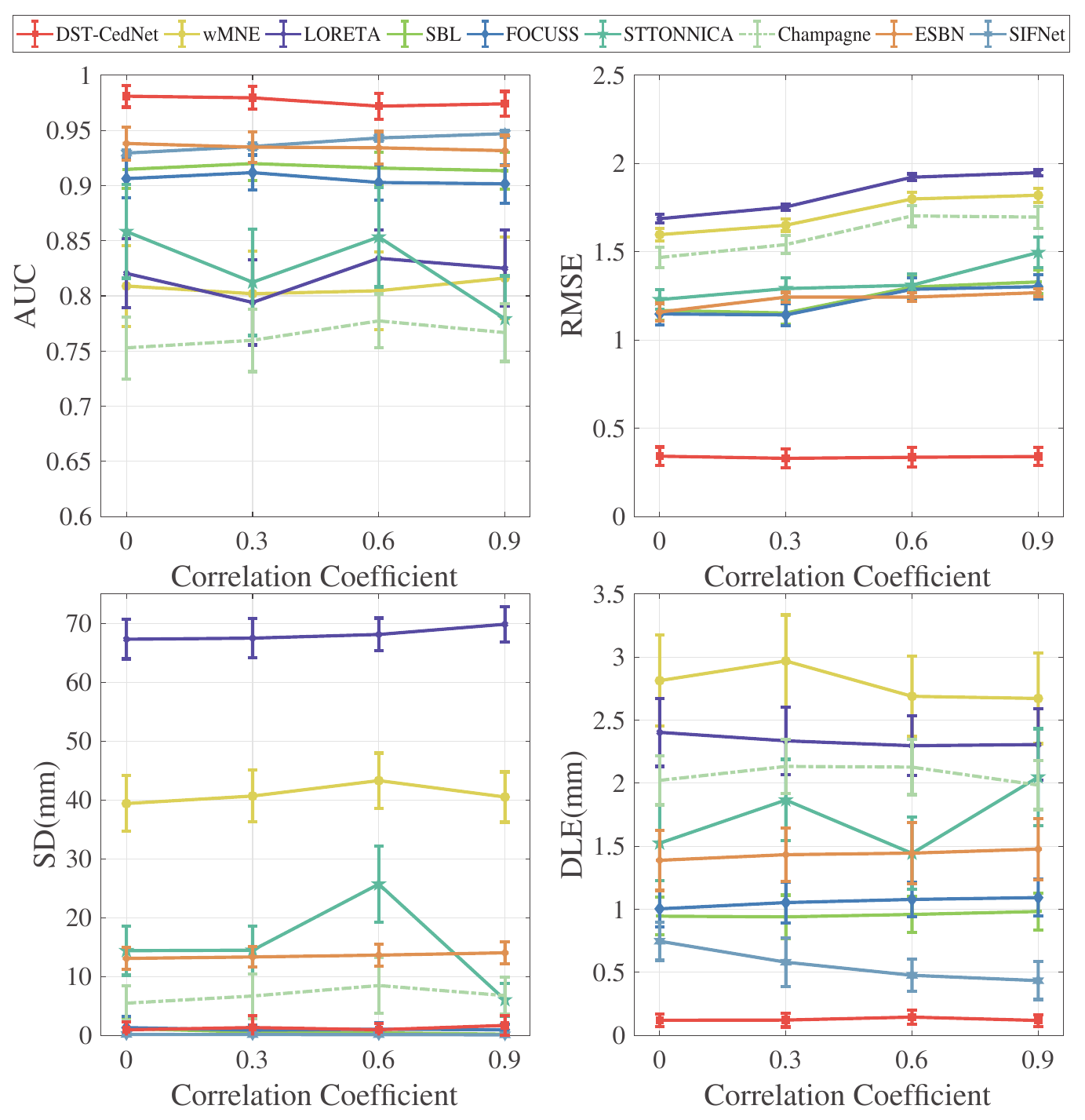}
		\caption{Performance metrics with varying correlated sources. Results are assessed across 100 Monte Carlo simulations and shown as mean $\pm$ SEM. SNR is -5 dB. Areas of the sources are all 7 $cm^2$. }
		%	\hfil
		\label{fig_10}
	\end{figure}
	
	\begin{figure*}[htbp]
		\centering
		\includegraphics[width=5.3in]{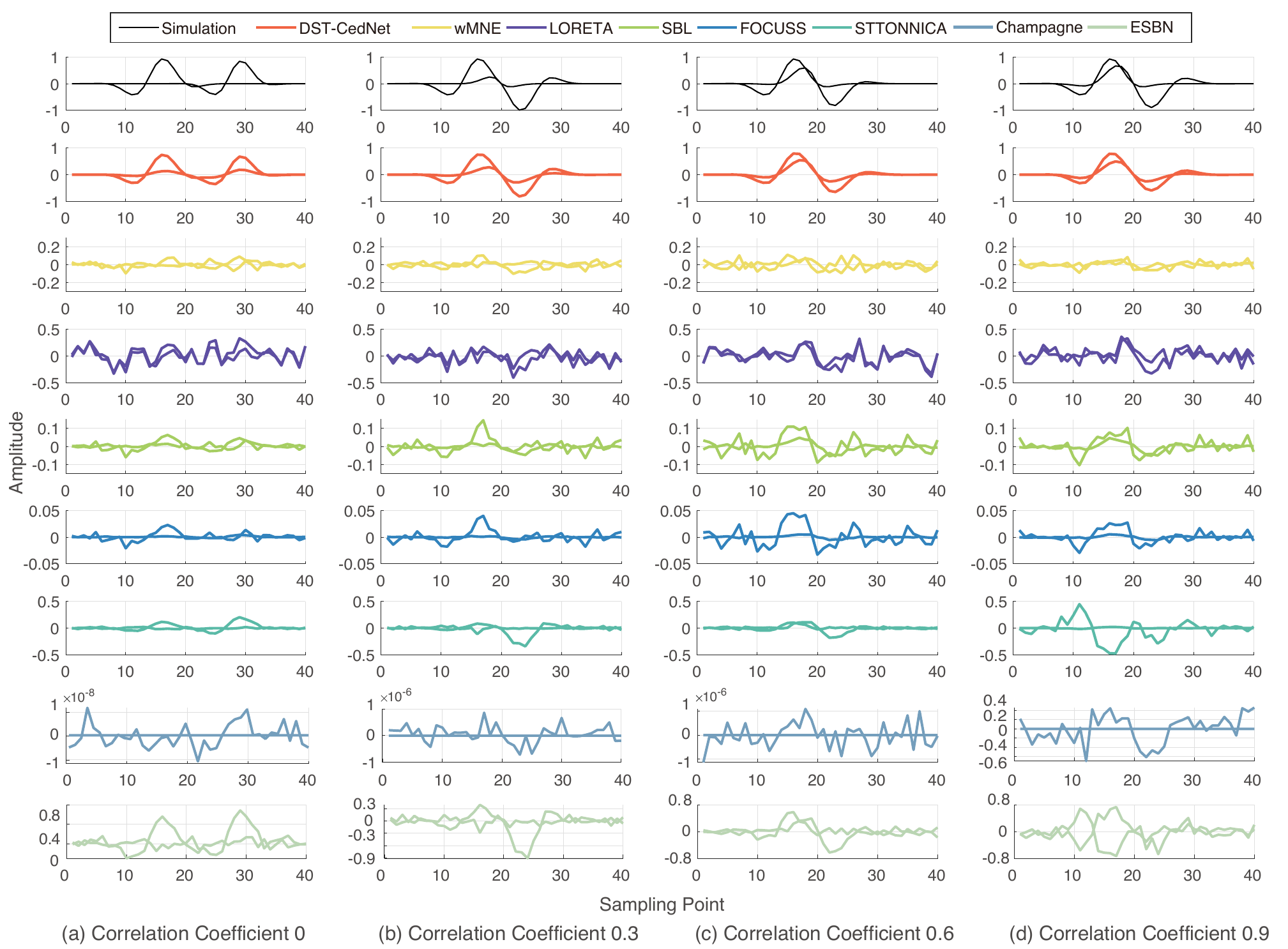}
		\caption{Reconstruction of time courses with varying correlated sources. The center of each source is selected as a reference for the comparison of different methods. The correlation coefficient between the time courses of the two center sources is varied. Note the range of the y-axis is adapted for each algorithm for visualization purpose. SNR is -5 dB.}
		\label{fig_11}
	\end{figure*}
	
	% Lastly we designed 4 groups of simulation sets with different correlated sources for the experiment. To explore the capability of the temporal reconstruction under multiple sources,
	
	Finally, we examine the performance of different algorithms when the activities of the two patch sources are correlated, where the areas of both sources are fixed to 7 $cm^2$ and the strength of the correlation is varied from 0 (uncorrelated) to 0.9 (highly-correlated).
	% modified in 7/3

% 	As shown in Fig. \ref{fig_10}, \textbf{under most settings DST-CedNet still dominates the performance metrics that remain stable as the source correlation varies (as well as other deep learning methods)}. By contrast, the metrics of other traditional algorithms are more susceptible to the change of the source correlation. In particular, among all the algorithms, STTONNICA shows the most dramatic fluctuations across all the metrics, with the worst performance at $r = 0.9$.

	As shown in Fig. \ref{fig_10}, \textbf{DST-CedNet achieves the best performance under most settings. As other deep learning methods, DST-CedNet's performance remains stable as the source correlation varies}. By contrast, the metrics of other traditional algorithms are more susceptible to the change of the source correlation. In particular, among all the algorithms, STTONNICA shows the most dramatic fluctuations across all the metrics, with the worst performance at $r = 0.9$.
	
	% which indicates that the proposed model has the superiority to distinguish similar temporal states. 
	% SBL, FOCUSS, or STTONNICA are shown to be relatively sensitive to the temporal state as the SD and DLE values fluctuate. STTONNICA performs the worst among them, showing drastic fluctuation on highly correlated sources. 
	
	% modified in 7/3
	To further assess the accuracy of estimating the source temporal dynamics, Fig. \ref{fig_11} shows the time courses of the reconstructed sources by different algorithms. The respective center source from each of the two correlated sources is selected as a reference for the comparison of different methods. \textbf{Among the compared algorithms, wMNE and LORETA fail to capture the temporal dynamics of the sources entirely. SBL, FOCUSS, Champagne, and STTONNICA are able to identify the peaks and troughs of one of the sources but not of the other. ESBN can recover both peaks of correlated sources but the temporal dynamics are noisy compared to the ground truth.} 
	% precise description 
	Note that the reconstructed waveforms by STTONNICA are relatively smooth due to its model assumption regarding the spatio-temporal decomposition that has incorporated temporal constraints. 
	As a consequence of the temporal smoothness of the training data and the powerful learning capability of the convolutional encoder-decoder, DST-CedNet can precisely recover temporal dynamics of both sources.

	\begin{figure}[htbp]
		\centering
		\includegraphics[width=3.5in]{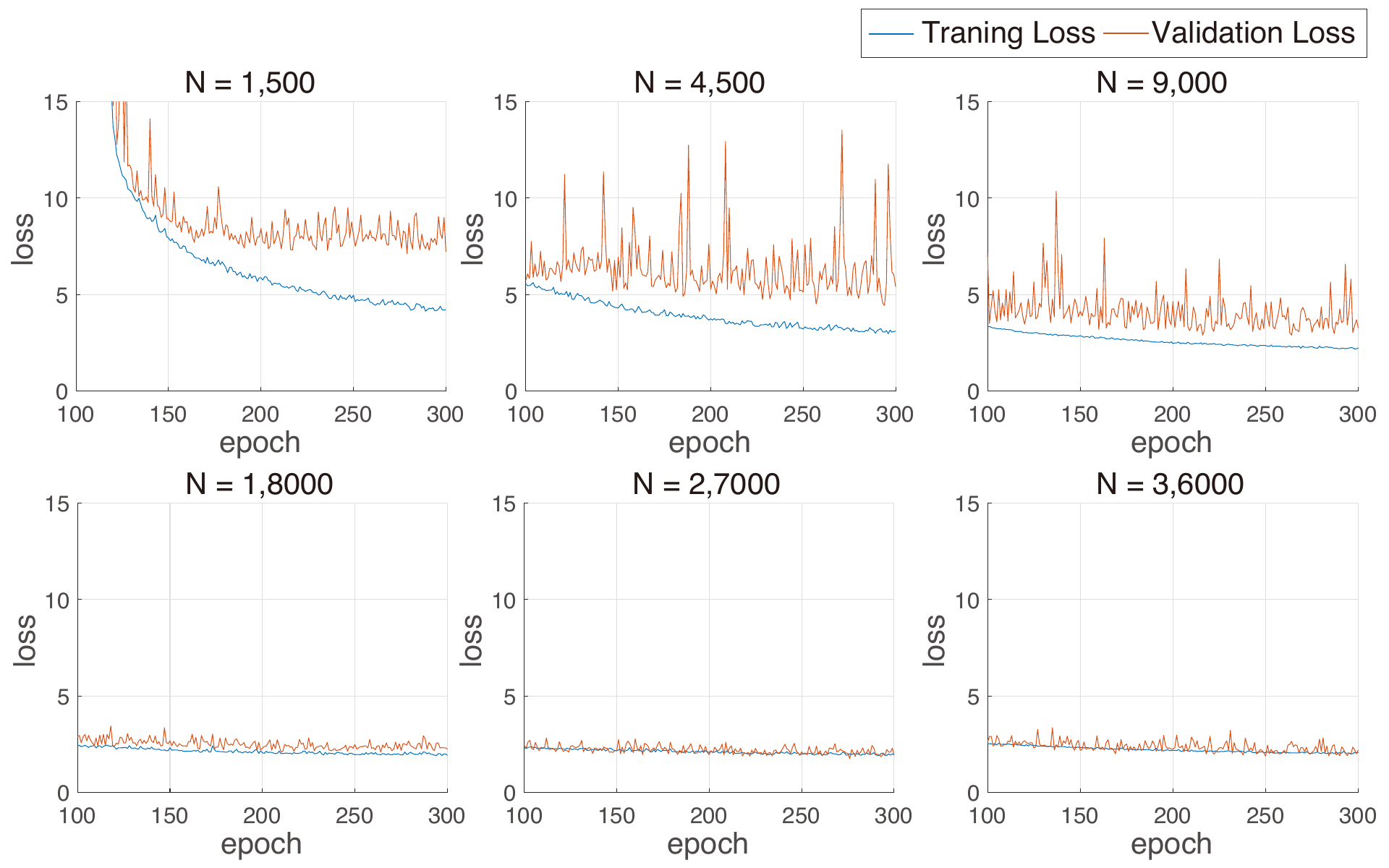}
		\caption{Training loss and validation loss with varying training sample size. Batch-size of the stochastic gradient descent is set to 32 and the epoch number of optimization to 250. $N$ is the training sample size.}
		%	\hfil
		\label{fig_13}
	\end{figure}
	\subsection{Impact of Training Sample Size} \label{sec.experiments.PNN}
	
	% modified in 6/7
	\begin{table}[htbp] 	
		\setlength\tabcolsep{5pt}
		\caption{$R^2$ metric of source signals with varying training sample size. $R^2$ is measured between the simulated and estimated source signals in the validation set.}
		\begin{tabular}{ccccccc}
			
			\toprule
			Training Sample & 1,500 & 4,500 & 9,000 & 18,000& 27,000 & 36,000\\
			\midrule
			$R^2$ & 0.253 & 0.448 & 0.686 & 0.856 & 0.913 & 0.922 \\
			
			\bottomrule
		\end{tabular}

		\label{Tab_4}
	\end{table}

	% modified in 6/7: test set --> validation set
	% modified in 6/22: rephrase
    To study the impact of the training sample size on the performance of DST-CedNet, we train the neural network with data of different sample sizes under the single patch source condition. We progressively increases the sample size from 1,500 to 36,000, with the batch size of the stochastic gradient descent fixed at 32 and the epoch number during optimization set to 250. As shown in Fig. \ref{fig_13}, the training and validation loss functions as well as their gap decrease as the training sample size increases from 1,500 to 27,000 where a plateau is reached, suggesting the gradual improvement of both the representation and generalization performance of the neural network. This observation is corroborated by the trend of the $R^2$ score shown in Table \ref{Tab_4}, where the increase is small when the training sample size reaches 27,000.

	\subsection{Experiment on Real MEG Data} \label{sec.experiments.MEG}

	Having validated DST-CedNet on the simulated data, we further investigate its performance in the analysis of a real MEG data set\footnote{http://neuroimage.usc.edu/bst/download.php}. The MEG median nerve data of one subject were collected from a stimulation experiment on the left and right thumb, which include 301 and 302 trials, respectively. The experiment was performed at Massachusetts General Hospital using the Neuromag Vectorview 306 system. There are a total of 306 recording sensors, including 102 MEG magnetometers and 204 MEG planar gradiometers, for which the sampling rate is 1,793 Hz. The length of each trial is 400 ms, including a 100 ms prestimulus baseline. Here we use the 301 left-hand trials for analysis. Moreover, following the suggested data preprocessing settings\footnote{https://neuroimage.usc.edu/brainstorm/Tutorials/TutMindNeuromag\ttvar{#}Pre-processing}, we select the 102 magnetometer channels and data within the poststimulus 250 ms time window (starting from 0 ms) as the test signal for DST-CedNet. The lead field matrix is calculated based on the subject's individual anatomy via the overlapping spheres model\cite{huang1999sensor}. 
	%Fig. \ref{fig_17} shows the pre-stimulus and post-stimulus scalp signals.

	The TBFs of the MEG data are estimated via SEFA, which represent a total of 80$\%$ of the variance of the signal. To reduce the number of training parameters, we ultimately obtain 48 sampling points after antialiasing filtering and downsampling. The cortex space is down-sampled to 1,024 to facilitate the calculation of the convolution kernel. The number of spatial components $K$ is set to 3. The SNR of the noise corruption is estimated using prestimulus baseline data. The batch size for SGD is set at 32. The Adam optimizer is selected, in which the learning rate is set to $10^{-4}$, hyperparameter $\beta_1$ is 0.9, $\beta_2$ is 0.999, and $\epsilon$ is $10^{-8}$. The number of training epochs is set to 250.
	
\begin{figure}[htbp]
		\centering
		\includegraphics[width=0.475\textwidth]{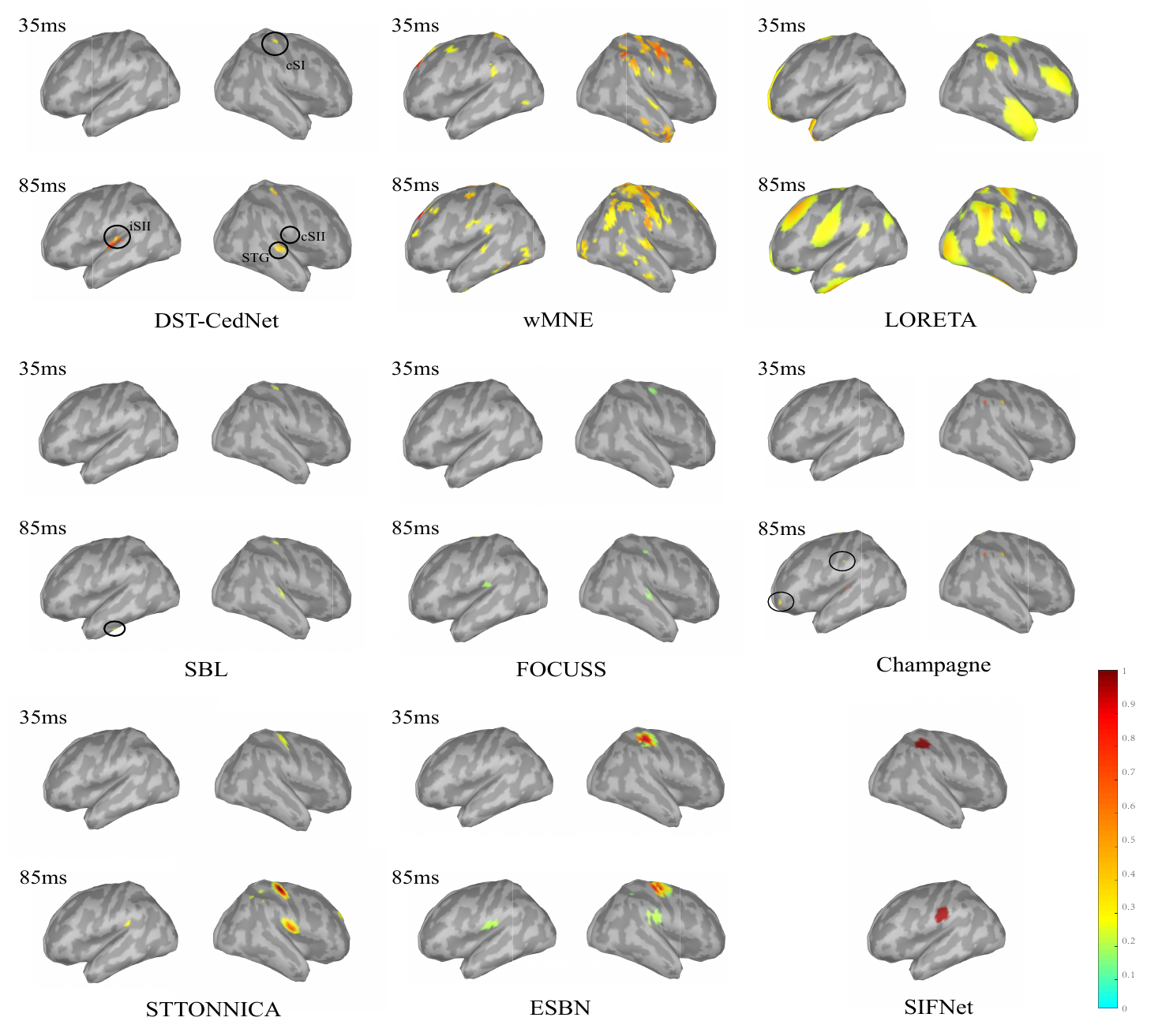}
		\caption{Comparison of source imaging of MEG data at 35 and 85 ms. The MEG median nerve data was collected in a stimulation experiment on the thumbs. The contralateral primary somatosensory cortex (cSI) area, the secondary somatosensory cortices (SII) area, and the superior temporal gyrus (STG) area are circled in the first row.}
		%	\hfil
		\label{fig_14}
	\end{figure}

    Fig. \ref{fig_14} depicts the source imaging results at 35 ms and 85 ms, as previous studies\cite{hari1999magnetoencephalography} reported that there are substantial brain activities in the contralateral primary somatosensory cortex (cSI) area at 35 ms, and in the cSI area and the secondary somatosensory cortex (SII) area at 85 ms. These prior findings are served as the ``ground truth" for comparing different algorithms. \textbf{As can be observed, the reconstructed sources by wMNE and LORETA are diffuse in which the activation is smeared into the motor area. SBL, FOCCUS, and Champagne yield highly focal sources. However, iSII and cSII are not accurately localized by SBL and Champagne, respectively; spurious sources appear in the left ventral temporal cortex for FOCCUS and Champagne at 85 ms. As for STTONNICA, although it can locate the cSI area at 35 ms as well as cSI and contralateral SII (cSII) at 85 ms, it does not accurately localize the ipsilateral SII (iSII) area at 85 ms but instead yields a few spurious sources in the vicinity of cSI. ESBN is able to localize both the cSI and SII areas. However, spurious sources are noticeable at 85 ms. The cSII area is entirely missed by SIFNet, which is unable to resolve temporal information since it only aims to classify source states.}
	% modified in 7/5
	Interestingly, it is noted that the superior temporal gyrus (STG) region is also activated at 85 ms in the reconstructed source map by DST-CedNet. This finding is consistent with the previous source imaging studies \cite{ou2009distributed} where STG rather than cSII is typically activated in the reconstructed maps. This may be explained by the fact that STG and cSII are structurally closely connected.

	\subsection{Experiment on Real EEG Data} \label{sec.experiments.EEG}
\begin{figure}[htbp]
		\centering
		\includegraphics[width=0.475\textwidth]{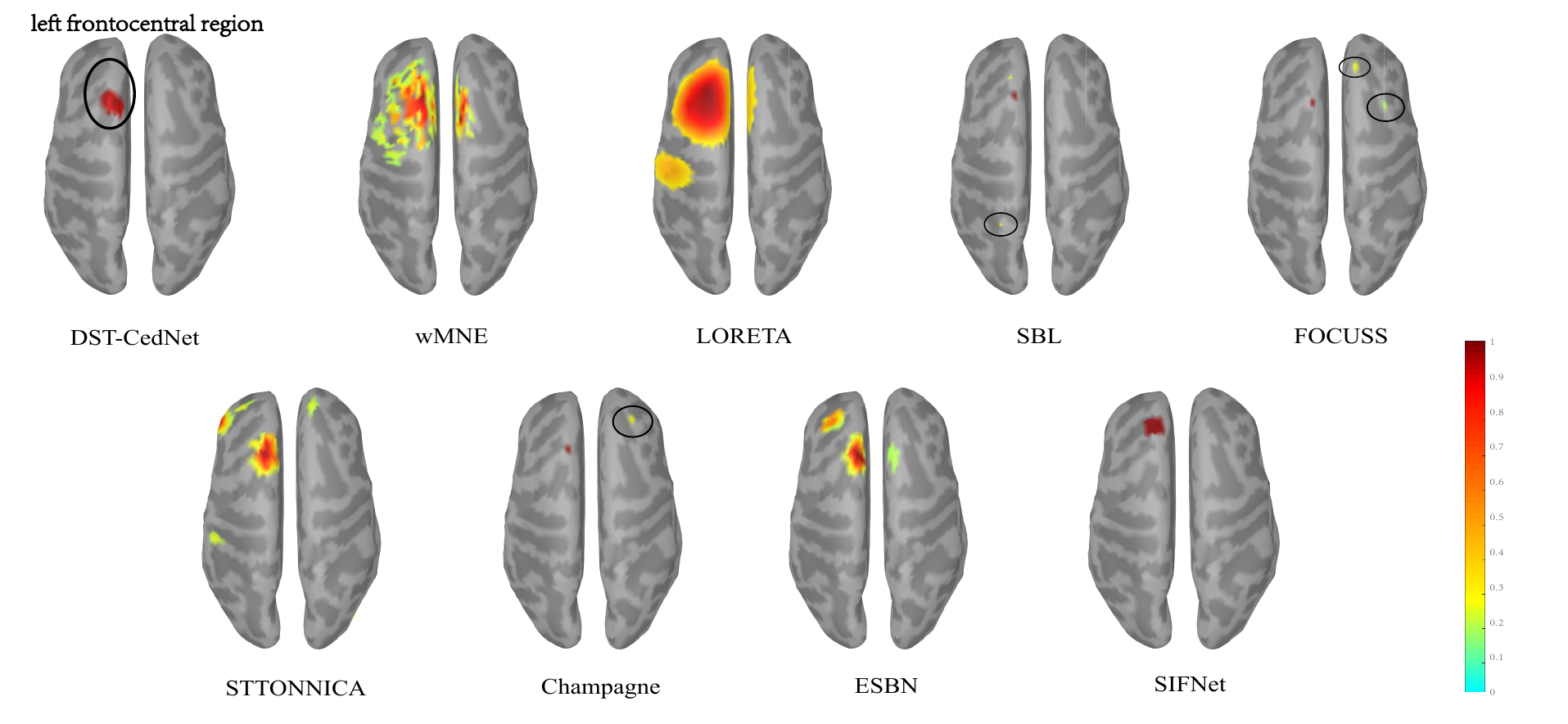}
		\caption{Comparison of source imaging of Epilepsy EEG data at 0 ms. The data was collected in a patient who suffered from left frontal epilepsy. Source activities are occurred in the left frontocentral region as reported by a previous study with invasive EEG recordings\cite{dumpelmann2012sloreta}.}
		%	\hfil
		\label{fig_15}
	\end{figure}

    \textbf{To further assess DST-CedNet's performance on real-world data, a public epilepsy EEG dataset\footnote{https://neuroimage.usc.edu/brainstorm/DatasetEpilepsy} is used. 
    There are a total of 41 recording electrodes, including 29 EEG electrodes and reference electrodes, for which the sampling rate is 256 Hz. The length of each trial is 1000 ms, with a 300 ms prestimulus baseline. 
    The average spikes, head model and the lead field matrix are computed by following the procedures outlined in the Brainstorm tutorial. 
    The EEG data was recorded from a patient suffering from focal frontal epilepsy, who underwent a left frontal tailored resection and was seizure-free with a follow-up of 5 years. Invasive EEG recordings on the same patient [52] suggested that source activities occurred in the left frontocentral region. Fig. \ref{fig_15} depicts the reconstructed sources using different methods. As can be seen, all methods are able to locate the sources in the left frontal lobe. Nonetheless, wMNE and LORETA yield sources that are spatially too smooth, while the sources obtained by SBL, FOCUSS, and Champagne are too focal. STTONNICA and ESBN are able to reconstruct sources with good spatial contiguity and homogeneity in the amplitude, but spurious sources are also observable in other regions. The sources estimated by SIFNet are located more towards the prefrontal region. In line with the invasive finding, DST-CedNet more accurately localizes focal sources in the frontocentral region without spurious activities.
    }
    
	% discuss
	\section{Discussion}
	\label{sec.discussion}
	
	% the contribution of the proposed method
    In this paper, we propose a novel data-synthesis-based deep learning approach to solve the ESI problem. 
	 % comparison
	Compared with LORETA, wMNE, SBL, FOCCUSS, STTONNICA, Champagne, ESBN, and SIFNet, DST-CedNet achieves higher detection sensitivity and smaller error in estimating the source extent and shape in simulation experiments. In addition, DST-CedNet yields more neurophyisiologically plausible source estimation on a realistic MEG and EEG dataset. 
	 % superiority 
	The superiority of our method is owing to how the prior information is leveraged in the algorithm. Specifically, a notable difference between DST-CedNet and traditional ESI methods is that the prior information regarding source signals is not enforced via mathematical constraints but rather the data synthesis strategy that generates a variety of states from the spatial-temporal prior space, which further guarantees the generalizability of the neural network and the validity of the simulated data.
	 % details
	  % spatial
	Previous studies suggested that brain activities tend to be spatially contiguous and locally homogeneous\cite{hamalainen1993,tao2005,destexhe1999}. Our method exploits these spatial properties to generate spatial states, which enables the network to yield estimates with compact and accurate extents. 
		  % temporal
	Unlike traditional methods\cite{zumer2008,trujillo2008} where TBFs are directly utilized as the temporal components of source signals under probabilistic graphic models, our method instead uses TBFs as the basis to generate the temporal components of the source signals, which is more likely able to capture the diverse dynamics of brain activities.	
		% modified in 7/6: modified the subsection.
	\textbf{Moreover, DST-CedNet can robustly extract spatio-temporal features from highly noisy E/MEG signals due to the denoising capability via the latent-space representation learning task introduced in the CedNet. The latent-space representation provided by the CedNet allows DST-CedNet to extract high-level features from noisy E/MEG signals, which effectively regularizes the solution obtained from discriminative learning}.

    \subsection{Generalization Capability}
    \textbf{
	To investigate the generalization capability of our approaches, we examine the performance of the model trained with a specific SNR (-5 dB or 10 dB) on validation sets with different SNRs. Performance metrics, namely $R^2$, AUC, RMSE, SD, and DLE, are then calculated on the validation set with an equal size as the training set. As shown in Table \ref{tab1}, the model trained with a -5 dB SNR generalizes well to validation sets with varying SNRs in terms of $R^2$, with moderate performance degradation as the SNR in the validate sets increases. By contrast, the model trained with 10 dB SNR generalizes poorly to validation sets with a -5 dB SNR. This result demonstrates that good generalization performance can be achieved with our model trained with low SNR datasets.}
    
    \begin{table*}[htbp]
	\centering
	\caption{Performance metrics on a single patch source with varying SNRs in the validation set. Results are assessed across 100 Monte Carlo simulations and shown as mean $\pm$ SEM. Models are trained with an SNR of -5 dB and 10 dB, respectively. }
\resizebox{\textwidth}{!}
{
	\begin{tabular}{ccccccccc}
		\toprule                                                                              \\ 
		Training SNR (dB) & \multicolumn{4}{c}{-5 ($R^2$ = 0.7392)}    & \multicolumn{4}{c}{10 ($R^2$ = 0.7805))} \\ 
		 \cmidrule(r){2-5} \cmidrule(r){6-9}  
		Test SNR (dB) & -5 & 0 & 5 & 10  & -5 & 0 & 5 & 10 \\ 
		\midrule
		$R^2$   &$\bm{0.7145}\pm$0.010 &$\bm{0.6651}\pm$ 0.015 &$\bm{0.6268}\pm$0.019 &$\bm{0.5839}\pm$ 0.022  
				&$\bm{0.3420}\pm$0.031 &$\bm{0.5378}\pm$0.024 &$\bm{0.6928}\pm$0.012  &$\bm{0.7234}\pm$0.008 
				\\
				
        AUC     &0.9819$\pm$0.0151 &0.9716$\pm$0.0152  &0.9677$\pm$0.0219 &0.9559$\pm$0.0286 
                &0.9493$\pm$0.0181 &0.9568$\pm$0.0121 &0.9781$\pm$0.0088 &0.9846$\pm$0.0033
				\\
	
        RMSE    &0.3893$\pm$0.1342 &0.4314$\pm$0.1450 &0.5486$\pm$0.2290 &0.5793$\pm$0.2313
                &0.8209$\pm$0.1701 &0.5199$\pm$0.1513 &0.3590$\pm$0.1482 &0.3274$\pm$0.1479
				\\
				
	   SD     & 0.0826$\pm$0.0589 &0.0913$\pm$0.0854 &0.1964$\pm$0.1603 &0.5005$\pm$0.3903 
            & 0.4753$\pm$0.3525  &0.3013$\pm$0.2682 &0.1042$\pm$0.1986 &0.0152$\pm$0.0097
            \\        
            
        DLE    &0.1463$\pm$0.0632 &0.2283$\pm$0.1016 &0.3883$\pm$0.1769 &0.4875$\pm$0.2458
            &0.4384$\pm$0.1604 &0.2422$\pm$0.0992 &0.1852$\pm$0.0909 & 0.1228$\pm$0.0607    
            \\		

		\bottomrule
	\end{tabular}\label{tab1}
}
\end{table*}
    
	\subsection{ Hyperparameter Tuning } \label{sec.discussion.HT}

	In the proposed model, key hyperparameters are the kernel size, the stride (step size), and the number of feature maps. Here we assess the impact of these hyperparameters on the predictive performance of the model. We first generate a training set and a validation set with equal sample size ($N = 13,500$ trials) using our data synthesis approach. Each trial of 64-channel synthetic signals contains 40 time points and is corrupted by an additive noise with an SNR of -5 dB. The source space consists of 1,024 vertices. Subsequently, we train multiple models by varying the hyperparameters using the training set, where the training batch size is 32 and the number of epochs is 250. The predictive performance of the trained models is then compared on the validation set. Since hyperparameter search is time-consuming, we only search within a confined set of hyperparameter values in order to save computational cost. Moreover, each time we only vary one hyperparameter with respect to a base model. The kernel size of the base model is $1\times3$, $1\times4$, $8\times1$, and $10\times1$, for the temporal encoding blocks, temporal decoding blocks, spatial encoding blocks, and spatial decoding blocks, respectively. These kernel sizes are smaller than the ones in Section \ref{sec.methodology} for the purpose of reducing the computational cost. The stride is $1\times2$ and $2\times1$ for the temporal and spatial blocks, respectively. Finally, to account for potential variance of the data, we repeat the above procedure for 5 runs and calculated the average $R^2$ between the predicted source signals and the true source signals.

	\begin{table*}[htbp]
		\setlength\tabcolsep{5.5pt}
		\centering
		\caption{$R^2$ scores for source reconstruction with varying kernel sizes. Each column shows the performance of neural network with only the kernel size in the respective layer changed.}
		\begin{tabular}{ccccccccccccc}
			\toprule
			~       & \multicolumn{12}{c}{Kernel Size Settings}                                                                                   \\ 
			\cmidrule{2-13} 
			$R^2$ Score & \multicolumn{4}{c}{$Temporal$ $En/Decoder$}    & \multicolumn{4}{c}{$Spatial$ $Decoder$}    & \multicolumn{4}{c}{$Spatial$ $Encoder$}    \\ 
			\cmidrule(r){2-5} \cmidrule(r){6-9} \cmidrule{10-13} 
			~       &  1$\times$3/4   & 1$\times$5/6  & 1$\times$7/8 & 1$\times$9/10 & 20$\times$1 & 10$\times$1 & 6$\times$1 & 3$\times$1 &64$\times$1 & 32$\times$1  & 16$\times$1 & 8$\times$1 \\ 
			\midrule
			Training Set       & 0.738  & 0.734    & 0.726  & 0.697  & 0.768    & 0.736   & 0.711  & 0.678  & 0.774 & 0.756 & 0.749 & 0.735       \\
			Validation Set     & 0.717   & 0.721    & 0.701   & 0.681  & 0.697    & 0.715   & 0.689  & 0.650  & 0.732 & 0.729 & 0.723 & 0.714          \\
	%		Test Set        & 0.718   & 0.723    & 0.699  & 0.684  & 0.697    & 0.716  & 0.689    & 0.650   & 0.734 & 0.731 & 0.724 & 0.717        \\
			\midrule
			Total Parameters   & 312,250  & 335,226  & 358,202 & 381,178  & 394,170 & 312,250 & 279,482  & 254,906  & 417,978 & 451,130 & 377,786 & 312,250 \\
			\bottomrule
		\end{tabular}\label{Tab_6}
	\end{table*}

	\begin{table*}[htbp]
		\setlength\tabcolsep{5.5pt}
		\centering
		\caption{$R^2$ scores for source reconstruction with varying hyperparameter settings. Each column shows the performance of neural network with only the hyperparameter in the respective layer changed. TE, TD, and SD represent the temporal encoder, temporal decoder, and the spatial decoder, respectively. ``inc" and ``dec" stand for ``increase" and ``decrease", respectively.}
		\begin{tabular}{ccccccccccccc}
			\toprule
			~       & \multicolumn{12}{c}{Hyper-parameter Settings }\\
			
			\cmidrule{2-13} 
			$R^2$ Score & \multicolumn{3}{c}{$Step$ $Size$ $(TE)$}  & \multicolumn{3}{c}{$Step$ $Size$ $(TD)$} & \multicolumn{3}{c}{$Step$ $Size$ $(SD)$ }  & \multicolumn{3}{c}{$Feature$ $Map$ $(Encoder)$} \\
			
			\cmidrule(r){2-4} \cmidrule(r){5-7} \cmidrule(r){8-10} \cmidrule(r){11-13}   
			~       &  2   & 4  & 8 &  2   & 4  & 8 &  2   & 4  & 8 & 2x inc     & constant  & 2x dec  \\ 
			
			\midrule
			Training Set       & 0.737   & 0.733  & 0.620  & 0.736   & 0.725  & 0.488 & 0.738   & 0.747  & 0.757 & 0.736  & 0.716   & 0.719  \\
			Validation Set     & 0.720   & 0.706  & 0.585 & 0.718   & 0.716  & 0.467 & 0.719   & 0.714  & 0.712  & 0.718  & 0.683  & 0.689  \\
			%		Test Set        & 0.720  & 0.704  & 0.364      & 0.718  & 0.686    & 0.690  & 0.718 & 0.687           \\
			\midrule
			Total Parameters  & 312,250  & 305,146 & 300,586 & 312,250  & 307,066 & 300,242 & 312,250  & 281,242 & 270,906 & 312,250 & 155,970 & 330,130 \\
			\bottomrule
		\end{tabular}\label{Tab_7}
	\end{table*}

	\subsubsection{Kernel Size} The kernel size determines the receptive field of the neural network and hence influences the modeling capability – the larger the kernel size is, the more global features the neural network is able to capture. However, the model complexity will also increase with the expansion of the parameter space.

	Table \ref{Tab_6} shows the predictive performance of the neural network by varying the kernel size, where each column corresponds to varying the kernel size of one block while fixing the other hyperparameters in the base model. For the temporal blocks, a temporal encoder kernel of $1\times5$ and a temporal decoder kernel of $1\times6$ lead to the highest $R^2$ score in the validation set. Moreover, the difference of the predictive performance between the training set and the validation set is smaller than those associated with other kernel sizes. For the spatial decoder, the optimal kernel size is $20\times1$. Notably, the predictive performance declines dramatically when the kernel size increases to 8. In terms of the spatial encoder, a kernel size equal to the channel number achieves the best predictive performance, which is in line with the EEG decoding results reported in \cite{schirrmeister1703deep}.

	\subsubsection{Step Size} The step size (i.e., stride) of the kernel refers to the rate of downsampling in the decoder layers and upsampling in the encoder layers, which determines the scale of the local information each feature map can capture and affects the number of the neural network layers. For instance, an input with 40 dimensions needs 3 layers with the stride of 2 to be reduced to an output with $40/2^3 = 5$ dimensions. We separately test three different strides, namely 2, 4, and 8, for the temporal encoder and decoder and the spatial decoder. As shown in Table \ref{Tab_7}, the predictive performance of the neural network deteriorates as the step size of the temporal encoder/ decoder increases. Notably, a drastic decrease of the performance occurs when the step size is 8 since a step size significantly larger than the kernel size may lead to adverse effects on training the neural network. Besides, the overfitting issue aggravates as the step size of the spatial decoder increases.

	\subsubsection{Feature Map} We compare three modes for the number of feature maps: 1) the number of feature maps increases for the encoder and decreases for the decoder by 2-fold between adjacent layers; 2) the number of feature maps is fixed to 32 for the encoder while decreases by 2-fold between adjacent decoder layers; and 3) the number of feature maps decreases for both the encoder and decoder by 2-fold between adjacent layers. As shown in Table \ref{Tab_7}, the first mode attains the best predictive performance, arguing for an hourglass-shaped structure for the proposed neural network.

	\subsection{Time Complexity}
	For the real MEG dataset, the total number of trainable parameters is 509,846, the batch size of the stochastic gradient descent is 32, and the number of training epochs is 250. With the network architecture and hyperparameters fixed, the time complexity depends on the number of training samples, which is $\mathcal{O}(n)$. The mean run-times per epoch for different training sample sizes on a single 12 GB TITAN-Xp GPU are shown in Table \ref{Tab_5}. Furthermore, to measure the time it takes to make predictions using the trained neural network, we made 100 predictions on the real MEG data that contains 48 sampling points across 102 channels. The average prediction runtime for each dataset is 3.598$\pm$0.152 sec.
	
	\begin{table}[htbp] 	
		\setlength\tabcolsep{4pt}
		\caption{ Mean runtimes per epoch for different training sample sizes on a single 12 GB TITAN-Xp GPU. The number of training epochs is 250.}
		\begin{tabular}{lcccccc}
			\toprule
			Training sample size & 1,500 & 4,500 & 9,000 & 18,000& 27,000 & 36,000\\
			\midrule
			Time/epoch (s) & 9.2 & 28.8 & 61.2& 125.2& 170.2 & 232.4 \\
			\bottomrule
		\end{tabular}

		\label{Tab_5}
	\end{table}
	
In DST-CedNet, although the training process is time-consuming, once trained the model can be applied to new data to make predictions without the need of optimization as traditional methods do.

	\subsection{Limitations and Future Directions}
    Reconstructing a high-resolution source space entails the estimation of a large number of parameters and hyperparameters prohibitive to the available computing resource. In order to ease the computational burden, we reduce the dimensionality of the parameter space by sampling the source space at a relatively low spatial resolution. This limitation is expected to become less problematic with the rapid growth in computing power.  

	% the decline in multiple sources 
	In this paper, we have focused on assessing the predictive performance of DST-CedNet for single patch and two patch sources, where the number of the patch source is assumed to be known in generating the training data. It remains to be tested whether the predictive performance persists when the number of patch sources exceeds two. Since more source configurations may lead to highly similar E/MEG patterns in the case of multiple patch sources, a potential challenge will be designing an appropriate loss function sensitive to the subtle difference between these source configurations. \textbf{Additionally, in this study we were unable to perform quantitative evaluation of our method on real data since the precise locations of the true sources are unavailable for both public datasets considered in this paper. Future studies will seek opportunities to validate our method using concurrent intracranial and scalp EEG recordings with precise coordinate information.}
	
	%Moreover, according to Section \ref{sec.experiments.PNN}, the predictive performance of DST-DAE is decided by the network architecture and loss function rather than the scale of the training data beyond a certain sample size. 
	% limitation in other deep learning methods 
	Moreover, to the best of our knowledge, current deep learning-based ESI methods restrict the possible numbers of patch sources in a single training set, as mixing different numbers of patch sources in the training set may lead to a decrease of the reconstruction performance. Therefore, our future work will focus on the design of better architectures and loss functions for dealing with multisources. 
 
 	As a final note,  although the encoding and decoding blocks in our network structure are tailored to E/MEG signals, it remains a challenge to interpret the features in the intermediate layers in terms of their neurophysiological meaning and how they impact source reconstruction, as is often the case for deep neural networks due to their black-box nature. Such model interpretability is crucial for clinical and health care practice, and will be studied in our future work.

	% conc
	\section{Conclusion}\label{sec.conclusion}
	% TODO: modified the conclusion
	In conclusion, we have proposed a novel neural network method termed DST-CedNet to solve the ESI problem. Compared with traditional methods that rely on optimization constraints to estimate sources, DST-CedNet generate large-scale training sets via a data synthesis strategy, allowing the neural network to automatically learn the sensor-to-source mapping. By leveraging prior knowledge regarding the spatio-temporal characteristics of brain activities during data generation, prior information can be readily incorporated without the need of sophisticated mathematical modeling and can conveniently be applied to new data without further optimization. Furthermore, the neural network integrates discriminative learning and latent-space representation under the framework of the Convolutional encoder-decoder, \textbf{where the auxiliary latent-space representation task in the CedNet designed with specific blocks is able to extract a robust spatio-temporal feature from highly noisy E/MEG signals, effectively regularizing the solution obtained from discriminative learning.} Both simulation and realistic data analyses demonstrate that DST-CedNet outperforms traditional and cutting edge deep learning ESI approaches in a range of source settings. Therefore, the method provides a useful tool for source localization in both basic science and translational studies.

\section*{Acknowledgment}\label{sec.acknowledgment}
The authors would like to thank Chen Wei and Kexin Lou at Shenzhen Key Laboratory of Smart Healthcare Engineering, Department of Biomedical Engineering, Southern University of Science and Technology for assisting with the implementation of the ESBN algorithm.

	% reference
	\bibliographystyle{IEEEtran}
	\bibliography{reference}

\begin{thebibliography}{10}
\providecommand{\url}[1]{#1}
\csname url@samestyle\endcsname
\providecommand{\newblock}{\relax}
\providecommand{\bibinfo}[2]{#2}
\providecommand{\BIBentrySTDinterwordspacing}{\spaceskip=0pt\relax}
\providecommand{\BIBentryALTinterwordstretchfactor}{4}
\providecommand{\BIBentryALTinterwordspacing}{\spaceskip=\fontdimen2\font plus
\BIBentryALTinterwordstretchfactor\fontdimen3\font minus
  \fontdimen4\font\relax}
\providecommand{\BIBforeignlanguage}[2]{{%
\expandafter\ifx\csname l@#1\endcsname\relax
\typeout{** WARNING: IEEEtran.bst: No hyphenation pattern has been}%
\typeout{** loaded for the language `#1'. Using the pattern for}%
\typeout{** the default language instead.}%
\else
\language=\csname l@#1\endcsname
\fi
#2}}
\providecommand{\BIBdecl}{\relax}
\BIBdecl

\bibitem{BinHe2011}
B.~He, L.~Yang, C.~Wilke, and H.~Yuan, ``Electrophysiological imaging of brain
  activity and connectivity{\textemdash}challenges and opportunities,''
  \emph{{IEEE} Transactions on Biomedical Engineering}, vol.~58, no.~7, pp.
  1918--1931, Jul. 2011.

\bibitem{Michel2004}
C.~M. Michel, M.~M. Murray, G.~Lantz, S.~Gonzalez, L.~Spinelli, and R.~G.
  de~Peralta, ``{EEG} source imaging,'' \emph{Clinical Neurophysiology}, vol.
  115, no.~10, pp. 2195--2222, Oct. 2004.

\bibitem{Baillet2001}
S.~Baillet, J.~Mosher, and R.~Leahy, ``Electromagnetic brain mapping,''
  \emph{{IEEE} Signal Processing Magazine}, vol.~18, no.~6, pp. 14--30, 2001.

\bibitem{Hmlinen1994}
M.~S. H\"{a}m\"{a}l\"{a}inen and R.~J. Ilmoniemi, ``Interpreting magnetic
  fields of the brain: minimum norm estimates,'' \emph{Medical {\&} Biological
  Engineering {\&} Computing}, vol.~32, no.~1, pp. 35--42, Jan. 1994.

\bibitem{Dale1993}
A.~M. Dale and M.~I. Sereno, ``Improved localizadon of cortical activity by
  combining {EEG} and {MEG} with {MRI} cortical surface reconstruction: A
  linear approach,'' \emph{Journal of Cognitive Neuroscience}, vol.~5, no.~2,
  pp. 162--176, Apr. 1993.

\bibitem{PascualMarqui1994}
R.~Pascual-Marqui, C.~Michel, and D.~Lehmann, ``Low resolution electromagnetic
  tomography: a new method for localizing electrical activity in the brain,''
  \emph{International Journal of Psychophysiology}, vol.~18, no.~1, pp. 49--65,
  Oct. 1994.

\bibitem{Grech2008}
R.~Grech, T.~Cassar, J.~Muscat, K.~P. Camilleri, S.~G. Fabri, M.~Zervakis,
  P.~Xanthopoulos, V.~Sakkalis, and B.~Vanrumste, ``Review on solving the
  inverse problem in {EEG} source analysis,'' \emph{Journal of
  {NeuroEngineering} and Rehabilitation}, vol.~5, no.~1, p.~25, 2008.

\bibitem{Becker2015}
H.~Becker, L.~Albera, P.~Comon, R.~Gribonval, F.~Wendling, and I.~Merlet,
  ``Brain-source imaging: From sparse to tensor models,'' \emph{{IEEE} Signal
  Processing Magazine}, vol.~32, no.~6, pp. 100--112, Nov. 2015.

\bibitem{Gorodnitsky1995}
I.~F. Gorodnitsky, J.~S. George, and B.~D. Rao, ``Neuromagnetic source imaging
  with {FOCUSS}: a recursive weighted minimum norm algorithm,''
  \emph{Electroencephalography and Clinical Neurophysiology}, vol.~95, no.~4,
  pp. 231--251, Oct. 1995.

\bibitem{ding2008sparse}
L.~Ding and B.~He, ``Sparse source imaging in electroencephalography with
  accurate field modeling,'' \emph{Human brain mapping}, vol.~29, no.~9, pp.
  1053--1067, 2008.

\bibitem{xu2007lp}
P.~Xu, Y.~Tian, H.~Chen, and D.~Yao, ``Lp norm iterative sparse solution for
  eeg source localization,'' \emph{IEEE transactions on biomedical
  engineering}, vol.~54, no.~3, pp. 400--409, 2007.

\bibitem{Wipf2009}
D.~Wipf and S.~Nagarajan, ``A unified bayesian framework for {MEG}/{EEG} source
  imaging,'' \emph{{NeuroImage}}, vol.~44, no.~3, pp. 947--966, Feb. 2009.

\bibitem{owen2012performance}
J.~P. Owen, D.~P. Wipf, H.~T. Attias, K.~Sekihara, and S.~S. Nagarajan,
  ``Performance evaluation of the champagne source reconstruction algorithm on
  simulated and real m/eeg data,'' \emph{Neuroimage}, vol.~60, no.~1, pp.
  305--323, 2012.

\bibitem{wu2015bayesian}
W.~Wu, S.~Nagarajan, and Z.~Chen, ``Bayesian machine learning:
  Eeg$\backslash$/meg signal processing measurements,'' \emph{IEEE Signal
  Processing Magazine}, vol.~33, no.~1, pp. 14--36, 2015.

\bibitem{Wipf2010}
D.~P. Wipf, J.~P. Owen, H.~T. Attias, K.~Sekihara, and S.~S. Nagarajan,
  ``Robust bayesian estimation of the location, orientation, and time course of
  multiple correlated neural sources using {MEG},'' \emph{{NeuroImage}},
  vol.~49, no.~1, pp. 641--655, Jan. 2010.

\bibitem{cai2020robust}
C.~Cai, M.~Diwakar, A.~Hashemi, S.~Haufe, K.~Sekihara, and S.~S. Nagarajan,
  ``Robust estimation of noise for electromagnetic brain imaging with the
  champagne algorithm,'' \emph{NeuroImage}, p. 117411, 2020.

\bibitem{liu2015straps}
K.~Liu, Z.~L. Yu, W.~Wu, Z.~Gu, and Y.~Li, ``Straps: A fully data-driven
  spatio-temporally regularized algorithm for m/eeg patch source imaging,''
  \emph{International journal of neural systems}, vol.~25, no.~04, p. 1550016,
  2015.

\bibitem{ValdsSosa2009}
P.~A. Vald{\'{e}}s-Sosa, M.~Vega-Hern{\'{a}}ndez, J.~M. S{\'{a}}nchez-Bornot,
  E.~Mart{\'{\i}}nez-Montes, and M.~A. Bobes, ``{EEG} source imaging with
  spatio-temporal tomographic nonnegative independent component analysis,''
  \emph{Human Brain Mapping}, vol.~30, no.~6, pp. 1898--1910, Jun. 2009.

\bibitem{Awan2018}
F.~G. Awan, O.~Saleem, and A.~Kiran, ``Recent trends and advances in solving
  the inverse problem for {EEG} source localization,'' \emph{Inverse Problems
  in Science and Engineering}, vol.~27, no.~11, pp. 1521--1536, Jul. 2018.

\bibitem{Knsche2013}
T.~R. Kn\"{o}sche, M.~Gr\"{a}ser, and A.~Anwander, ``Prior knowledge on cortex
  organization in the reconstruction of source current densities from {EEG},''
  \emph{{NeuroImage}}, vol.~67, pp. 7--24, Feb. 2013.

\bibitem{Dale2000}
A.~M. Dale, A.~K. Liu, B.~R. Fischl, R.~L. Buckner, J.~W. Belliveau, J.~D.
  Lewine, and E.~Halgren, ``Dynamic statistical parametric mapping,''
  \emph{Neuron}, vol.~26, no.~1, pp. 55--67, Apr. 2000.

\bibitem{arridge2019solving}
S.~Arridge, P.~Maass, O.~{\"O}ktem, and C.-B. Sch{\"o}nlieb, ``Solving inverse
  problems using data-driven models,'' \emph{Acta Numerica}, vol.~28, pp.
  1--174, 2019.

\bibitem{abeyratne1991artificial}
U.~R. Abeyratne, Y.~Kinouchi, H.~Oki, J.~Okada, F.~Shichijo, and K.~Matsumoto,
  ``Artificial neural networks for source localization in the human brain,''
  \emph{Brain Topography}, vol.~4, no.~1, pp. 3--21, 1991.

\bibitem{van2000eeg}
G.~Van~Hoey, J.~De~Clercq, B.~Vanrumste, R.~Van~de Walle, I.~Lemahieu,
  M.~D'Hav{\'e}, and P.~Boon, ``Eeg dipole source localization using artificial
  neural networks,'' \emph{Physics in Medicine \& Biology}, vol.~45, no.~4, p.
  997, 2000.

\bibitem{yuasa1998eeg}
M.~Yuasa, Q.~Zhang, H.~Nagashino, and Y.~Kinouchi, ``Eeg source localization
  for two dipoles by neural networks,'' in \emph{Proceedings of the 20th Annual
  International Conference of the IEEE Engineering in Medicine and Biology
  Society. Vol. 20 Biomedical Engineering Towards the Year 2000 and Beyond
  (Cat. No. 98CH36286)}, vol.~4.\hskip 1em plus 0.5em minus 0.4em\relax IEEE,
  1998, pp. 2190--2192.

\bibitem{abeyratne2001eeg}
U.~R. Abeyratne, G.~Zhang, and P.~Saratchandran, ``Eeg source localization: a
  comparative study of classical and neural network methods,''
  \emph{International journal of neural systems}, vol.~11, no.~04, pp.
  349--359, 2001.

\bibitem{goodfellow2016deep}
I.~Goodfellow, Y.~Bengio, and A.~Courville, \emph{Deep learning}.\hskip 1em
  plus 0.5em minus 0.4em\relax MIT press, 2016.

\bibitem{Sun2020sifnet}
R.~Sun, A.~Sohrabpour, S.~Ye, and B.~He, ``Sifnet: Electromagnetic source
  imaging framework using deep neural networks,'' \emph{bioRxiv}, 2020.

\bibitem{wei2021edge}
C.~Wei, K.~Lou, Z.~Wang, M.~Zhao, D.~Mantini, and Q.~Liu, ``Edge sparse basis
  network: A deep learning framework for eeg source localization,'' in
  \emph{2021 International Joint Conference on Neural Networks (IJCNN)}, 2021,
  pp. 1--8.

\bibitem{He1987}
B.~He, T.~Musha, Y.~Okamoto, S.~Homma, Y.~Nakajima, and T.~Sato, ``Electric
  dipole tracing in the brain by means of the boundary element method and its
  accuracy,'' \emph{{IEEE} Transactions on Biomedical Engineering}, vol.
  {BME}-34, no.~6, pp. 406--414, Jun. 1987.

\bibitem{Hamalainen1989}
M.~Hamalainen and J.~Sarvas, ``Realistic conductivity geometry model of the
  human head for interpretation of neuromagnetic data,'' \emph{{IEEE}
  Transactions on Biomedical Engineering}, vol.~36, no.~2, pp. 165--171, Feb.
  1989.

\bibitem{schirrmeister1703deep}
R.~T. Schirrmeister, J.~T. Springenberg, L.~D.~J. Fiederer, M.~Glasstetter,
  K.~Eggensperger, M.~Tangermann, F.~Hutter, W.~Burgard, and T.~Ball, ``Deep
  learning with convolutional neural networks for brain mapping and decoding of
  movement-related information from the human eeg,'' \emph{arXiv preprint
  arXiv:1703.05051}, 2017.

\bibitem{8310961}
S.~{Sakhavi}, C.~{Guan}, and S.~{Yan}, ``Learning temporal information for
  brain-computer interface using convolutional neural networks,'' \emph{IEEE
  Transactions on Neural Networks and Learning Systems}, vol.~29, no.~11, pp.
  5619--5629, 2018.

\bibitem{8897723}
O.~{Kwon}, M.~{Lee}, C.~{Guan}, and S.~{Lee}, ``Subject-independent
  brain-computer interfaces based on deep convolutional neural networks,''
  \emph{IEEE Transactions on Neural Networks and Learning Systems}, pp. 1--14,
  2019.

\bibitem{dumoulin2016guide}
V.~Dumoulin and F.~Visin, ``A guide to convolution arithmetic for deep
  learning,'' \emph{arXiv preprint arXiv:1603.07285}, 2016.

\bibitem{odena2016deconvolution}
A.~Odena, V.~Dumoulin, and C.~Olah, ``Deconvolution and checkerboard
  artifacts,'' \emph{Distill}, vol.~1, no.~10, p.~e3, 2016.

\bibitem{clevert2015fast}
D.-A. Clevert, T.~Unterthiner, and S.~Hochreiter, ``Fast and accurate deep
  network learning by exponential linear units (elus),'' \emph{arXiv preprint
  arXiv:1511.07289}, 2015.

\bibitem{ioffe2015batch}
S.~Ioffe and C.~Szegedy, ``Batch normalization: Accelerating deep network
  training by reducing internal covariate shift,'' \emph{arXiv preprint
  arXiv:1502.03167}, 2015.

\bibitem{tao2005}
J.~X. Tao, A.~Ray, S.~Hawes-Ebersole, and J.~S. Ebersole, ``Intracranial eeg
  substrates of scalp eeg interictal spikes,'' \emph{Epilepsia}, vol.~46,
  no.~5, pp. 669--676, 2005.

\bibitem{hamalainen1993}
M.~H{\"a}m{\"a}l{\"a}inen, R.~Hari, R.~J. Ilmoniemi, J.~Knuutila, and O.~V.
  Lounasmaa, ``Magnetoencephalography—theory, instrumentation, and
  applications to noninvasive studies of the working human brain,''
  \emph{Reviews of modern Physics}, vol.~65, no.~2, p. 413, 1993.

\bibitem{destexhe1999}
A.~Destexhe, D.~Contreras, and M.~Steriade, ``Spatiotemporal analysis of local
  field potentials and unit discharges in cat cerebral cortex during natural
  wake and sleep states,'' \emph{Journal of Neuroscience}, vol.~19, no.~11, pp.
  4595--4608, 1999.

\bibitem{nagarajan2007probabilistic}
S.~S. Nagarajan, H.~T. Attias, K.~E. Hild, and K.~Sekihara, ``A probabilistic
  algorithm for robust interference suppression in bioelectromagnetic sensor
  data,'' \emph{Statistics in medicine}, vol.~26, no.~21, pp. 3886--3910, 2007.

\bibitem{ou2009distributed}
W.~Ou, M.~S. H{\"a}m{\"a}l{\"a}inen, and P.~Golland, ``A distributed
  spatio-temporal eeg/meg inverse solver,'' \emph{NeuroImage}, vol.~44, no.~3,
  pp. 932--946, 2009.

\bibitem{gramfort2010openmeeg}
A.~Gramfort, T.~Papadopoulo, E.~Olivi, and M.~Clerc, ``Openmeeg: opensource
  software for quasistatic bioelectromagnetics,'' \emph{Biomedical engineering
  online}, vol.~9, no.~1, p.~45, 2010.

\bibitem{tadel2011brainstorm}
F.~Tadel, S.~Baillet, J.~C. Mosher, D.~Pantazis, and R.~M. Leahy, ``Brainstorm:
  a user-friendly application for meg/eeg analysis,'' \emph{Computational
  intelligence and neuroscience}, vol. 2011, 2011.

\bibitem{kingma2014adam}
D.~P. Kingma and J.~Ba, ``Adam: A method for stochastic optimization,''
  \emph{arXiv preprint arXiv:1412.6980}, 2014.

\bibitem{gotmare2018closer}
A.~Gotmare, N.~S. Keskar, C.~Xiong, and R.~Socher, ``A closer look at deep
  learning heuristics: Learning rate restarts, warmup and distillation,''
  \emph{arXiv preprint arXiv:1810.13243}, 2018.

\bibitem{alexander2015beware}
D.~L. Alexander, A.~Tropsha, and D.~A. Winkler, ``Beware of r 2: simple,
  unambiguous assessment of the prediction accuracy of qsar and qspr models,''
  \emph{Journal of chemical information and modeling}, vol.~55, no.~7, pp.
  1316--1322, 2015.

\bibitem{otsu1979threshold}
N.~Otsu, ``A threshold selection method from gray-level histograms,''
  \emph{IEEE transactions on systems, man, and cybernetics}, vol.~9, no.~1, pp.
  62--66, 1979.

\bibitem{huang1999sensor}
M.~Huang, J.~C. Mosher, and R.~Leahy, ``A sensor-weighted overlapping-sphere
  head model and exhaustive head model comparison for meg,'' \emph{Physics in
  Medicine \& Biology}, vol.~44, no.~2, p. 423, 1999.

\bibitem{hari1999magnetoencephalography}
R.~Hari and N.~Forss, ``Magnetoencephalography in the study of human
  somatosensory cortical processing,'' \emph{Philosophical Transactions of the
  Royal Society of London. Series B: Biological Sciences}, vol. 354, no. 1387,
  pp. 1145--1154, 1999.

\bibitem{dumpelmann2012sloreta}
M.~D{\"u}mpelmann, T.~Ball, and A.~Schulze-Bonhage, ``sloreta allows reliable
  distributed source reconstruction based on subdural strip and grid
  recordings,'' \emph{Human brain mapping}, vol.~33, no.~5, pp. 1172--1188,
  2012.

\bibitem{zumer2008}
J.~M. Zumer, H.~T. Attias, K.~Sekihara, and S.~S. Nagarajan, ``Probabilistic
  algorithms for meg/eeg source reconstruction using temporal basis functions
  learned from data,'' \emph{NeuroImage}, vol.~41, no.~3, pp. 924--940, 2008.

\bibitem{trujillo2008}
N.~J. Trujillo-Barreto, E.~Aubert-V{\'a}zquez, and W.~D. Penny, ``Bayesian
  m/eeg source reconstruction with spatio-temporal priors,'' \emph{Neuroimage},
  vol.~39, no.~1, pp. 318--335, 2008.

\end{thebibliography}
	
	\clearpage %change page

\end{document}